\documentclass[preprint,12pt,authoryear]{elsarticle}
\usepackage{amssymb}
\usepackage{amsmath}
\usepackage{amsmath,amsfonts}
\usepackage{subcaption}
\usepackage{array}
\usepackage[caption=false,font=normalsize,labelfont=sf,textfont=sf]{subfig}
\usepackage{textcomp}
\usepackage{stfloats}
\usepackage{url}
\usepackage{verbatim}
\usepackage{graphicx}
\usepackage{adjustbox}
\usepackage{listings}
\usepackage{color}
\usepackage{graphicx}
\usepackage{tabularx}
\usepackage{multirow}
\usepackage{listings}
\usepackage{color}
\usepackage{xcolor}
\usepackage{soul}
\usepackage{regexpatch}
\usepackage{pdfpages}

\journal{Journal of Systems and Software}

\usepackage{tabularx}
\usepackage[linesnumbered,ruled]{algorithm2e}

\usepackage[table,xcdraw]{xcolor}
\usepackage{balance}
\usepackage{amsmath}
\usepackage{graphicx}
\usepackage{textcomp}
\usepackage{xcolor}
\usepackage{comment}
\usepackage{booktabs}
\usepackage{multirow}
\usepackage{xspace}
\usepackage{tcolorbox}
\usepackage{graphicx,subcaption} 
\usepackage{array}
\usepackage{ragged2e}
\usepackage{xcolor}
\usepackage{listings}
\usepackage{tikz}
\usetikzlibrary{decorations.pathreplacing,calc,shapes,positioning,tikzmark}

\usepackage{makecell}
\usepackage{enumitem}
\usepackage{subcaption}
\usepackage{tcolorbox}  
\usepackage{lipsum}     

\newcommand{\PreserveBackslash}[1]{\let\temp=\\#1\let\\=\temp}
\newcolumntype{C}[1]{>{\PreserveBackslash\centering}p{#1}}
\newcolumntype{R}[1]{>{\PreserveBackslash\raggedleft}p{#1}}
\newcolumntype{L}[1]{>{\PreserveBackslash\raggedright}p{#1}}

\def\BibTeX{{\rm B\kern-.05em{\sc i\kern-.025em b}\kern-.08em
    T\kern-.1667em\lower.7ex\hbox{E}\kern-.125emX}}
    
\newcommand{\pa}[1]{\noindent\textbf{#1}}
\newcommand{\spec}{NL-specification}
\newcommand{\specS}{NL-specification\xspace}
\newcommand{\specs}{NL-specifications\space}
\newcommand{\pseudocode}[1]{\texttt{#1}}

\definecolor{mainColor}{HTML}{000000}    \definecolor{subColor}{HTML}{004400} 
\newtcolorbox{boxFindings}{
    fontupper = \bf\color{mainColor}, 
    boxrule = 1.5pt,
    colframe = subColor,
    rounded corners,
    arc = 5pt   
}

\newtcolorbox[auto counter, number within=section]{promptBox}[2][]{colframe=blue!50!black, colback=blue!10, coltitle=black, 
    fonttitle=\bfseries, title=Text Box~\thetcbcounter: #2,#1}

\definecolor{keyword}{rgb}{0.58, 0.0, 0.83}
\definecolor{string}{rgb}{0.2, 0.6, 0.2}
\definecolor{comment}{rgb}{0.5, 0.5, 0.5}
\definecolor{extra}{rgb}{0.58, 0.0, 0.83}
\definecolor{correctMark}{rgb}{0.80, 0.99, 0.80}
\definecolor{wrongMark}{rgb}{0.99, 0.80, 0.80}
\definecolor{targetMark}{rgb}{0.80, 0.80, 0.80}
\definecolor{highlight}{rgb}{0.80, 0.90, 0.80}
\newcommand{\RQOne}{Can LLM-generated specifications boost LLM-based code translations? }
\newcommand{\RQTwo}{When LLM-generated specifications can boost code translation and when not?}
\newcommand{\RQThree}{What is the impact of specifications on the quality of translated code?}

\definecolor{darkgreen}{HTML}{00994d}

\begin{document}
\begin{frontmatter}

\title{Specification-Driven Code Translation Powered by Large Language Models: How Far Are We?}

\author[1]{Soumit Kanti Saha\fnref{equal}}
\ead{s_soumit@live.concordia.ca}

\author[1]{Fazle Rabbi\fnref{equal}}
\ead{fazle.rabbi@mail.concordia.ca}

\author[1]{Tri Minh Triet Pham}
\ead{p_triet@encs.concordia.ca}

\author[2]{Song Wang}
\ead{wangsong@yorku.ca}

\author[1]{Jinqiu Yang}
\ead{jinqiu.yang@concordia.ca}

\fntext[equal]{These authors contributed equally to this work.}

\affiliation[1]{organization={Computer Science and Software Engineering, Concordia University},
            city={Montreal},
            state={QC},
            country={Canada}}

\affiliation[2]{organization={Lassonde School of Engineering, York University},
            city={Toronto},
            country={Canada}}

\begin{abstract}
Large Language Models (LLMs) are increasingly being applied across various domains, including code-related tasks such as code translation. Previous studies have explored using LLMs for translating code between different programming languages. Since LLMs are more effective with natural language, {using natural language as an intermediate representation in code translation tasks is an intuitively appealing approach. However, whether this benefit is general or highly context-dependent remains unclear.} In this work, we investigate using \specS as an intermediate representation for code translation. We evaluate our method using three datasets, five popular programming languages, and 29 language pair permutations. Our results show that using \specS alone does not lead to performance improvements. However, when combined with source code, {it provides gains in certain language pairs (notably with Python and C++ as source languages), while offering no consistent improvement overall.} Besides analyzing the performance of code translation, we also investigate the quality of the translated code and provide insights into the issues present in the translated code.
\end{abstract}


\end{frontmatter}

\section{Introduction}
Migrating legacy codebases through code translation is a challenging yet necessary step in modernizing and adapting software systems to current technologies. The process typically begins with a comprehensive analysis of the original code's specifications, which provide a detailed blueprint of the code's intended functionality. These specifications, along with the legacy code itself, serve as critical anchors during the translation process, offering a clear reference for understanding the existing code’s logic and design patterns. By carefully examining the legacy code, key components, such as business rules, data structures, and algorithms, need to be identified, which must be accurately translated into the new language or framework.

For large software systems, manually migrating code is a time-consuming and error-prone process. 
Hence, there is a growing need for automated approaches to generate code to a target language from the existing codebase, such as from C to Rust. 
To this end, research efforts are made to automate code translation~\citep{lachaux2020unsupervised, yin2024rectifiercodetranslationcorrector, pan2023stelocoderdecoderonlyllmmultilanguage,achiam2023gpt, yang2024exploring, macedo2024intertrans, nitin2024spectraenhancingcodetranslation}.
Recently, large language models (LLMs) have demonstrated strong capabilities in both code generation and code translation.
On the generation side, modern LLMs can synthesize executable code directly from natural language descriptions with high accuracy \citep{nijkamp2022codegen, zheng2023codegeex, dubey2024llama, lozhkov2024starcoder}, achieving state-of-the-art performance on benchmarks such as \citet{macedo2024intertrans}.
These results suggest that LLMs are highly effective at mapping natural language semantics to correct program logic.

In parallel, prior work has explored LLM-based code translation across programming languages \citep{pan2024lost, bhattarai2024enhancing}.
Most existing approaches treat translation primarily as a syntax-preserving transformation task, relying almost exclusively on the source code as context.
For example, \citet{pan2024lost} translate code across multiple languages using GPT-4 with the full source program as input, while \citet{bhattarai2024enhancing} incorporate retrieval-based augmentation to assist translation.
While effective in many cases, these methods offer limited insight into whether explicitly modeling program intent can improve translation robustness, particularly for semantically complex code.

To better preserve semantics across languages, some studies introduce intermediate representations (IRs) for code translation.
Compiler-level IRs such as LLVM IR \citep{szafraniec2022code} enable structured, language-agnostic transformations, but are inherently low-level and lack support for many popular languages.
More recently, INTERTRANS \citep{macedo2024intertrans} explores the use of high-level programming languages as intermediate representations for LLM-based translation.
While promising, this approach remains constrained by the expressiveness and availability of suitable intermediate languages and does not explicitly separate program intent from implementation details.


{ Despite these advances, a key gap remains. Unlike compiler-level IRs, which are low-level and language-constrained, or programming-language IRs (INTERTRANS), which require a suitable intermediate language to exist, natural language specifications are universally applicable across all language pairs and are the modality in which program intent is most naturally expressed, and the primary training signal for LLMs. This positions \specS as a conceptually distinct intermediate representation whose effectiveness we empirically characterize across success and failure regimes, rather than assuming uniform benefit.}

Motivated by this observation, { we investigate when and how incorporating natural language specification (\spec) affects LLM-based code translation.} We hypothesize that \specS can (1) expose high-level intent that may be implicit or obscured in source code, (2) complement source code by clarifying semantics for complex control flow or logic, and (3) enable higher-quality target code by allowing generation to focus on intent rather than syntactic correspondence.
Unlike prior work, we systematically study translation using source code only, \specs only, and their combination, while limiting post-hoc fixes to compilation errors to better isolate the effect of the \spec.

Our study builds upon prior benchmarks and methodologies but differs in its focus on understanding \emph{when and why} \specS help code translation. At the same time, we acknowledge that \spec-based translation may be less effective when specifications are incomplete or ambiguous. The limitations are discussed in detail in Section~\ref{sec:limitations}.

{ The \specS generated by the LLMs corresponds to pseudocode, which combines simplified program statements with natural-language descriptions. Prior work by \citet{hindle2016naturalness} shows that software exhibits strong regularities and can be modeled as a form of natural language. Motivated by this observation, we treat LLM-generated \specS as a natural-language representation of program semantics, suitable for guiding code translation.}

In this study, we investigate the use of \specS as an intermediate representation for code translation. We generate \specS from source code using an LLM and evaluate two translation strategies: (1) relying exclusively on the generated \spec, and (2) a hybrid approach combining the original source code with \spec.

In this work, we answer the following three research questions (RQs). 
\begin{itemize}
    \item \textbf{RQ1: \RQOne}
    Based on evaluation results on three popular datasets (Avatar~\citep{ahmad-etal-2021-avatar}, CodeNet~\citep{puri2021codenet}, and HumanEval~\citep{chen2021evaluating}), we find that { leveraging \specs improves translation for certain source languages (Python and C++), but does not consistently outperform source-only translation when averaged across all language pairs.}
    \item \textbf{RQ2: \RQTwo} We study the use of LLM-generated \specS in code translation, focusing on the properties of the generated specifications and the challenges faced by current LLMs in producing accurate and complete representations of program semantics. Our results show that {LLM-generated \specS can improve translation correctness in specific cases, particularly for structurally complex programs where the specification is accurate.} However, the effectiveness of this approach is limited by specification inaccuracies, such as missing edge cases, ambiguous descriptions, and incomplete semantic coverage, which remain a key bottleneck for current LLMs.
    Last, we compare the quality of translated code among \citep{pan2024lost} and our two approaches that leverage \spec. We find that \spec-driven code translation tends to generate code with fewer quality warnings from SonarQube. We present insights into the types of issues encountered in the translated code, offering a deeper understanding of the challenges in cross-language code translation.\\
\end{itemize}

In this paper, our contributions are as follows:
{
\begin{enumerate} 
    \item We investigate the use of \specS as an intermediate representation for LLM-based code translation, evaluating both a \specS-only approach and a hybrid approach that combines \specS with source code. We find that using source code alone achieves the best average translation accuracy, while combining source code with \specS yields improved results for specific source languages.
    \item We conduct a detailed analysis of success and failure cases to understand when \specS helps or hinders code translation. Our analysis shows that \specS can effectively assist LLMs when it correctly captures program semantics, but inaccuracies in the generated \specS directly propagate to translation errors.
    \item We incorporate an iterative error-repair step and evaluate its impact on translation correctness. We find that fixing only compilation errors improves translation accuracy by an average of 8.5\% for the \specS-only approach and 6.1\% for the hybrid approach, highlighting the importance of post-translation repair.
\end{enumerate}
}
\textbf{Data Availability:} Complete code, API usage, dataset guidelines, and additional implementation details are available here\footnote{https://github.com/frabbisw/Code-Translation-using-NL-Specification/}. 


{ The remainder of this paper is organized as follows. Section~2 describes the experimental design, including dataset preparation, test case construction, and LLM selection. Section~3 details our methodology, covering the translation and repair processes as well as the SonarQube analysis setup. Section~4 presents the experimental results and addresses the research questions. Section~5 discusses the internal and external validity of the study. Section~6 reviews related work, and Section~7 concludes the paper and outlines directions for future work.}

\section{Experiment Design}

\begin{figure*}
    \centering
    \includegraphics[width=0.95\linewidth]{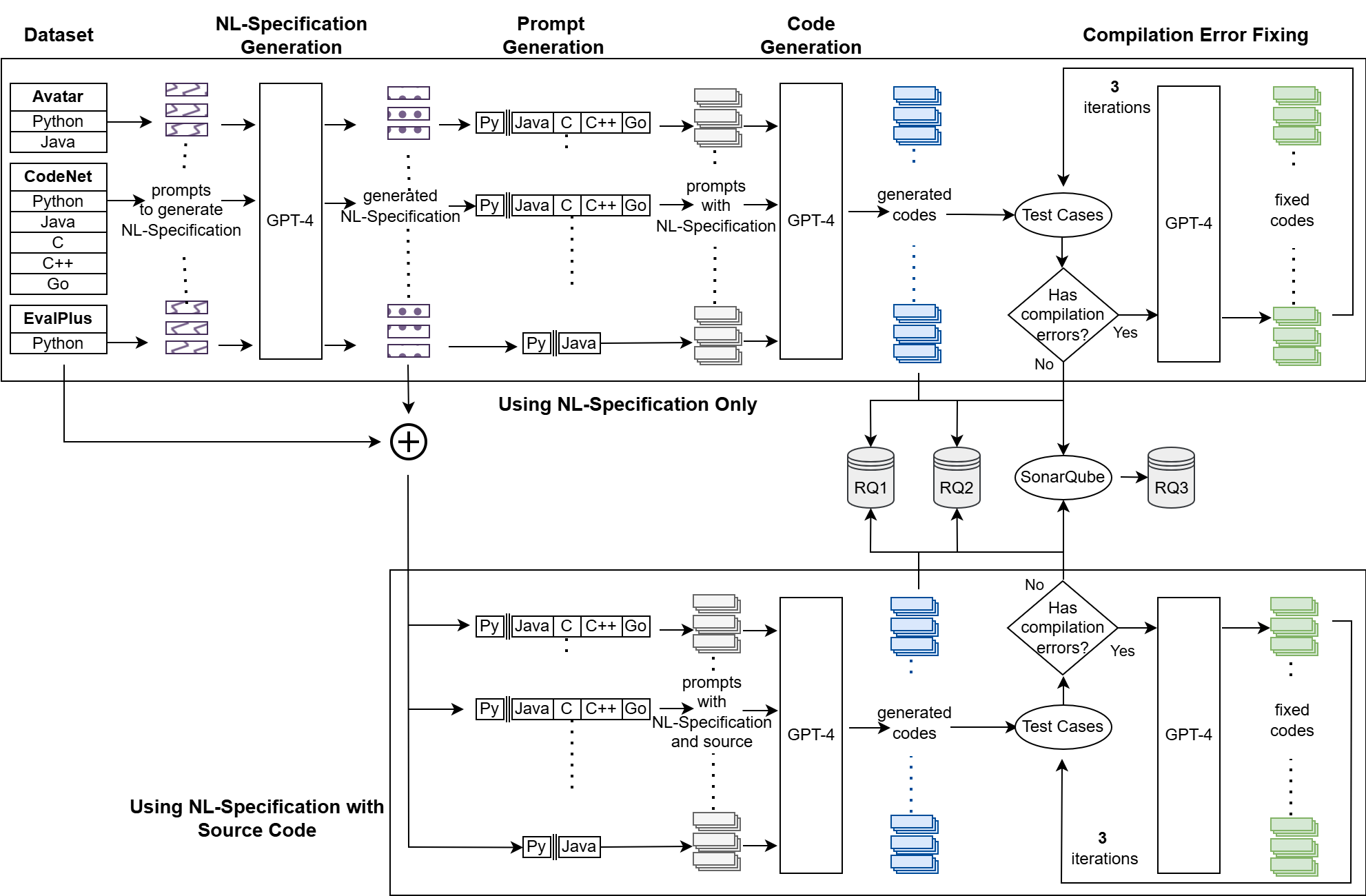}
    \caption{The workflow of our work.}
    \label{fig:overview}
\end{figure*}

\subsection{Dataset} In this section, the datasets used in this study are thoroughly discussed, along with the preparation carried out to make them ready for the experiments.
\subsubsection{Dataset Selection}
\label{sec:dataset}
{We use three datasets for our evaluation that have been widely used in prior works on code translation \citep{pan2024lost, nitin2024spectraenhancingcodetranslation, macedo2024intertrans} and generation ~\citep{chen2024supersonic,lozhkov2024starcoder, paul2024ircoder, dubey2024llama}. The datasets are CodeNet \citep{puri2021codenet}, Avatar \citep{ahmad-etal-2021-avatar}, and Evalplus \citep{liu2024your}.} HumanEval \citep{chen2021evaluating} and EvalPlus \citep{liu2024your} are the same datasets. However, EvalPlus contains more test cases.

The Avatar dataset contains a total of 9,515 Java and Python code snippets taken from programming contest sites such as Codeforces, AtCoder, etc. Among these, 250 instances come with test cases in the form of input and expected output. The CodeNet dataset, part of the IBM Project CodeNet initiative, is one of the largest and most diverse datasets available for AI research in code understanding and translation. It contains over 14 million code snippets across 50 programming languages and covers more than 4,000 different coding problems sourced from online coding competitions. Evalplus~\citep{liu2024your} is a benchmarking framework that extends HumanEval \citep{chen2021evaluating} and MBPP \citep{austin2021programsynthesislargelanguage} with extended unit tests. 
{Table~\ref{tab:dataset} summarizes the statistics of the three datasets.
Similar to prior works \citep{pan2024lost, yin2024rectifiercodetranslationcorrector, macedo2024intertrans}, we only select the code snippets that come with test cases. }

\begin{table}

\centering
\footnotesize
\caption{Statistics of experimental datasets. { CodeNet(1) is the subset used in LIT \citep{pan2024lost} and CodeNet(2) is the subset used in InterTrans \citep{macedo2024intertrans}}}

\begin{tabular}{|p{4cm}|p{2cm}|p{2cm}|p{2cm}|p{2cm}|}
\hline

\textbf{Dataset} & \textbf{Datapoints} & \textbf{No. of test cases} & \textbf{Min test case(s) per datapoint} & \textbf{Mean test case(s) per datapoint}\\
\hline

Avatar-Verified & 479 & 9381 & 1 & 19.6\\
CodeNet(1) & 1000 & 1000 & 1 & 1\\
EvalPlus & 164 & 2681 & 5 & 16.35\\
CodeNet(2) & 210 & 630 & 3 & 3\\
\hline
\end{tabular}
\label{tab:dataset}
\end{table}

\subsubsection{Dataset Validation}
{We successfully downloaded the CodeNet dataset. For Avatar, while the test cases were available, we were unable to download the corresponding code snippets by following the procedures described in the official GitHub repository \citep{ahmad-etal-2021-avatar}, as the download script referenced invalid links. As an alternative, we obtained the code snippets from the artifact released by \citet{pan2024lost}, which was evaluated and validated by ACM (version 1.1).

During validation, we identified multiple inconsistencies in the Avatar dataset where the expected outputs in the test cases did not match the actual outputs produced by the corresponding source code. In some cases, the expected outputs were incomplete and ended with ellipses (``...''). To address this, we executed the source code to obtain the actual outputs and compared them against the expected outputs using prefix matching, i.e., we checked whether the expected output was a prefix of the actual program output. When the prefix matched, we repaired the test case by replacing the expected output with the complete actual output.

For a small number of instances, no expected output matched the actual program output, making it impossible to construct valid test cases. We therefore excluded these instances from our experiments, consisting of 11 Python files and 10 Java files in Avatar. Overall, we verified and repaired incorrect outputs in the Avatar dataset to ensure test case validity. The dataset cleanup scripts are included in our artifact for transparency and reproducibility. We have named our cleaned Avatar Dataset as Avatar-Verified and published it in our repository. The results reported for the Avatar dataset from the baseline~\citep{pan2024lost} are based on the Avatar-Verified.}

Table \ref{tab:dataset} represents the number of valid instances and test cases after dataset cleanup. These are the final data we use for the whole experiment.

\subsubsection{Test Case Conversion}
{Since EvalPlus provides unit tests only for Python, we converted the Python test cases to other target languages (e.g., Java, C++, and JavaScript) to enable cross-language evaluation.

We follow the dataset setting of \citet{pan2024lost}, which builds on HumanEval-X and incorporates EvalPlus-style test cases for evaluation. Upon examining the released artifact of \citet{pan2024lost}, we observed that several unit tests were incomplete or not fully converted to Java. Specifically, some test files contained placeholder comments such as ``\textit{//\# Token size exceeded}'' or ``\textit{// Continue writing tests here following the format above}'', as found in the publicly released Zenodo artifact (v4, dated December 31, 2023).

To ensure consistent and valid evaluation across languages, we manually converted and completed the missing unit tests by translating the Python EvalPlus tests into the corresponding target-language formats. All converted test cases were validated by executing them against the canonical solutions. The conversion scripts and validated test cases are included in our artifact.}


{To ensure comprehensive validation of the EvalPlus dataset in Java, we developed a rule-based script to convert the original Python test cases into JUnit tests. The authors manually verified the converted tests to ensure correctness. The script also automatically updates parameter data types and corresponding assertions if the repair changes a parameter's type (for example, an $int[]$ can be transformed into an $ArrayList<Integer>$, and the JUnit test will be adapted accordingly).}

\subsection{Subject Programming Languages for Evaluation}

{We choose the same set of target programming languages as used in the baselines \citet{pan2024lost} and \citet{macedo2024intertrans} for comparison.} ~\citet{pan2024lost} selected the target programming languages according to their popularity (i.e., TIOBE index~\citep{tiobe}), covering different paradigms (procedural, functional, and object-oriented) and the availability of high-quality datasets.{ Thus, the selected programming languages were C, C++, Go, Java, and Python. The other baseline, \citet{macedo2024intertrans}, used C++, Go, Java, JavaScript, Python, and Rust.
However, all the languages that were selected as sources in the baselines are not available in Avatar and EvalPlus. In the Avatar dataset, only Java and Python are available as source programming languages, and the only available source programming language in the EvalPlus dataset is Python. 
}

\subsection{Subject LLMs}
\label{llms}
{To evaluate specification-driven code translation across diverse model families, we select a set of large language models that differ in architecture, training data, and accessibility. Following the methodology of \citet{pan2024lost}, which identifies OpenAI’s GPT models as strong performers in code translation, we include \textbf{GPT-4} as a representative closed-source model. GPT-4 is trained on a mixture of licensed data, data created by human trainers, and publicly available sources, with knowledge updated through October 2023. At the time of our experiments, it achieved state-of-the-art performance on code generation benchmarks (e.g., 90.2\% on HumanEval~\citep{chen2021evaluating}), making it well-suited for handling both code translation and natural-language specifications.

To assess the generalizability of our approach beyond proprietary models, we evaluate two additional strong \emph{open-source} code-specialized LLMs. \textbf{Magicoder-S-DS-6.7B}~\citep{wei2023magicoder} is a 6.7B-parameter model fine-tuned on instruction-style code data and is used in prior work on intermediate code translation, including \citet{macedo2024intertrans}, enabling direct comparison. \textbf{DeepSeek-Coder}~\citep{zhu2024deepseek} is a family of code-focused LLMs trained on large-scale code corpora across multiple programming languages, and has demonstrated strong performance on program synthesis and code understanding tasks.

By evaluating our approach using GPT-4 alongside Magicoder-S-DS-6.7B and DeepSeek-Coder, we demonstrate that our method is effective across both closed-source and open-source LLMs with varying model sizes and training strategies, providing evidence of its robustness and general applicability.}
{In total, the experiments in this study are conducted using three LLMs: GPT-4, Magicoder-S-DS-6.7B, and Deepseek-coder to demonstrate the generalizability of our method to a variety of open- and closed-sourced LLMs.}

\section{Methodology}
As shown in Fig. \ref{fig:overview}, we first take a code from our dataset and create the prompt 
to generate the \spec. Then, using the LLM, we generate the \specS for that corresponding code. 
We examined the correctness of the generated \spec. 
Then, we use two approaches to create the prompt. In the first approach, we create the prompt using only \spec. In the second approach, we create the prompt using source code and \specS to generate the target language code. Then we generate the target language code with the prompt. For a given datapoint, the \specS, the translated code, and the translation are each generated only once. {
Subsequently, the translated code is automatically compiled and validated using the dataset's test cases. Upon detecting a failure, such as a compilation error, functional mismatch, runtime exception, or non-termination, the pipeline constructs a repair prompt. This prompt aggregates the erroneous code with specific diagnostic feedback (such as compiler logs, stack traces, or failing test information) and is provided to the LLM for rectification. Thus, we try to generate a solution with error correction in an iterative approach. We stop iterating if the error is fixed or after three iterations. If there are no errors, we run the test files on the generated (or fixed) translation to evaluate the correctness.} We consider a translated code passed if all test cases are passed. In the following subsections, we describe each step in the pipeline, each accompanied by the prompt template.

\subsection{Generating \spec}

\begin{table}
\caption{Prompt template for generating \spec}
\vspace{-0.1in}
\begin{tabularx}{1\textwidth}{X}
\hline
\textbf{\textcolor{purple}{``prompt''}}: 
\textcolor{blue}{\{source\_code\}} \\
Give pseudocode for the above
\textcolor{blue}{\{source\_language\}} code so that the 
\textcolor{blue}{\{source\_language\}} code is reproducible from the pseudocode. Do not give any other explanation except for the pseudocode.
\\
\textbf{\textcolor{purple}{``output''}}: 
\textcolor{blue}{\{Answer\}} \\
\hline
\end{tabularx}
\label{tab:spec_gen}
\end{table}

According to our workflow as shown in Fig. \ref{fig:overview}, we first utilize an LLM to generate \specS from source code using the prompt (shown in table~\ref{tab:spec_gen}). The generated \specS is a line-by-line detailed NL description from the source code. An example of the generated \specS is shown in Fig. \ref{fig:sample_nl_spec}. From this figure (Generated \specS for HumanEval\_5.py), we can see that the func $intersperse$ takes a list of integers ($numbers$) and an integer ($delimeter$) and returns a list placing $delimeter$ in between all the consecutive integers of $numbers$. In the generated \specS the intersperse function is presented in natural language line by line from the source code also maintaining the code indentation. But does not contain the Python language-specific standard library imports. This \specS contains all the necessary details to write the same function in any language.

\subsection{Prompt Sensitivity Study}
\noindent We conducted a prompt sensitivity study on a representative dataset, EvalPlus, focusing on Python$\rightarrow$Java translation in the \specS setting. We evaluated three prompt variants for generating the intermediate representation:

\begin{itemize}
    \item \textbf{Pseudocode prompt}, which produces code-like, structured representations with explicit control flow and variable updates.  
    \\\emph{Example:} \\\texttt{Initialize sum = 0;\\for each element x in array: sum += abs(x);\\if negative\_count is odd, subtract 2 * min\_abs}

    \item \textbf{Structured natural-language specification prompt}, which provides a higher-level, abstract description of inputs, outputs, and algorithmic logic in prose.  
    \\\emph{Example:}\\\texttt{Compute the sum of absolute values of all numbers.\\If there is an odd number of negative values and no zeros,\\adjust the sum by subtracting twice the smallest absolute\\value.}

    \item \textbf{Control-flow–focused prompt}, which emphasizes execution order and decision structure without using language-specific syntax.  
    \\\emph{Example:}\\\texttt{Iterate through the list while tracking the number of\\ negative values and the smallest magnitude. After the loop,\\decide whether to adjust the final sum based on these\\conditions.}
\end{itemize}

All other factors, including the model (GPT-4), decoding parameters, and evaluation protocol, were held constant. {Across this subset, the pseudocode-based prompt achieved the highest compilation success and test pass rates. This is because pseudocode preserves the operational structure of the original program, statement order, variable updates, and loop bounds, which are precisely the details the LLM must reproduce in the target language. The more abstract NL variants tend to compress this structure, omitting state-update semantics that are critical for correctness on control-flow, heavy instances. Based on these results, we adopt the pseudocode prompt as the default \specS in the remainder of our experiments.}
Table~\ref{tab:prompt_ablation} reports the compilation rates and test pass rates for three different \specS prompt variants when translating from Python to Java on the EvalPlus dataset.

\begin{table}[h]

\centering
\footnotesize
\caption{Prompt sensitivity study on the Evalplus dataset
(Python$\rightarrow$Java, \specS translation).}
\label{tab:prompt_ablation}
\begin{tabular}{lcc}
\hline
\textbf{Prompt Variant} & \textbf{Comp. Error Rate (\%)} & \textbf{Test Pass Rate (\%)} \\
\hline
\textbf{(i) Pseudocode} & \textbf{3.66} & \textbf{77.44} \\
(ii) Structured NL Specification & 13.41 & 67.68 \\
(iii) Control-Flow Description & 10.98 & 73.17 \\
\hline
\end{tabular}
\end{table}

\begin{figure}
\centering
\tiny
\begin{tabular}{|m{\linewidth}|}
\hline
\textbf{\footnotesize Source Code $(HumanEval\_5.py)$} \\ \hline
\begin{lstlisting}[escapechar=~]
1 . from typing import List
2 . def intersperse(numbers: List[int], delimeter: int) 
3 . -> List[int]:
4 .      """ Insert a number 'delimeter' between every two  
5 . consecutive elements of input list `numbers'
6 .     >>> intersperse([], 4)
7 .     []
8 .     >>> intersperse([1, 2, 3], 4)
9 .     [1, 4, 2, 4, 3]
10.     """
11.     res = []
12.     for i in range(len(numbers)):
13.        res.append(numbers[i])
14.         if i != len(numbers) - 1:
15.             res.append(delimeter)
16.     return res
\end{lstlisting} \\ \hline
\textbf{\footnotesize Generated \spec} \\ \hline
\pseudocode{\tiny 1. FUNCTION intersperse(numbers: List of integers, delimeter: integer) -> List of integers}\\
\pseudocode{\tiny 2. \space\space INITIALIZE res as an empty list} \\
\pseudocode{\tiny 3. \space\space FOR i FROM 0 TO length of numbers - 1 DO} \\
\pseudocode{\tiny 4. \space\space\space\space APPEND numbers[i] to res} \\
\pseudocode{\tiny 5. \space\space\space\space IF i is NOT equal to length of numbers - 1 THEN} \\
\pseudocode{\tiny 6. \space\space\space\space\space\space APPEND delimeter to res}\\
\pseudocode{\tiny 7. \space\space END FOR}\\
\pseudocode{\tiny 8. \space\space RETURN res}\\
\pseudocode{\tiny 9. END FUNCTION}\\
\hline
\end{tabular}
\caption{An example of source code and generated \spec}
\label{fig:sample_nl_spec}
\end{figure}

\begin{table}
\caption{Prompt template for generating Target Language Code using only \spec}
\vspace{-0.1in}
\begin{tabularx}{1\textwidth}{X}
\hline
\textbf{\textcolor{purple}{``prompt''}}: 
\textcolor{blue}{\{pseudocode\_content\}} \\
The above pseudocode was generated from
\textcolor{blue}{\{source\_language\}}. Generate functionally
correct and similar \textcolor{blue}{\{target\_language\}}
code using the pseudocode. Print only the
\textcolor{blue}{\{target\_language\}} code and end with the
comment $\backslash$"End of Code$\backslash$". Do not give any
other explanation.\\
\textbf{\textcolor{purple}{``output''}}: 
\textcolor{blue}{\{Answer\}} \\
\hline
\end{tabularx}
\label{tab:target_gen_nl}
\end{table}

\begin{table}
\footnotesize
\caption{Prompt template for generating Target Language Code using source code and \spec}
\vspace{-0.1in}
\begin{tabularx}{1\textwidth}{X}
\hline
\textbf{\textcolor{purple}{``prompt''}}: 
\textcolor{blue}{\{source\_code\}} \\
This is a \textcolor{blue}{\{source\_language\}} code. \\ 
\textcolor{blue}{\{pseudocode\_content\}} \\
The above pseudocode was generated from
\textcolor{blue}{\{source\_language\}}. Generate functionally
correct and similar \textcolor{blue}{\{target\_language\}}
code using the pseudocode. Print only the
\textcolor{blue}{\{target\_language\}} code and end with the
comment $\backslash$``End of Code$\backslash$''. Do not give any
other explanation.\\
\textbf{\textcolor{purple}{``output''}}: 
\textcolor{blue}{\{Answer\}} \\
\hline
\end{tabularx}
\label{tab:target_gen_nl_source}
\end{table}

\subsection{Translating to target programming language}
\noindent We experimented with two ways to utilize \specs{} in code translation. First, only the generated \specs{} is used for code translation, and no source code is provided. Second, a combination of source code and generated \specs{} is provided for code translation.   
The prompts for the above-mentioned are shown in table \ref{tab:target_gen_nl} and table \ref{tab:target_gen_nl_source}. 

As shown in Table \ref{tab:target_gen_nl}, we have included the source language of the generated \specS to ensure correct API interpretation. Since distinct languages often share API names (e.g., min()) despite having different underlying behaviors and output structures, providing the language context is necessary for the LLM to process the native APIs correctly.

{We evaluated the correctness of all the generated code translations using the validation tests that come with the datasets. Similar to works in LLM code generation \citep{nijkamp2022codegen, dubey2024llama,lozhkov2024starcoder}, we calculate Computational Accuracy (CA) as the percentage of correct code transitions. 

\subsection{Fixing Errors in Translated Codes}
\begin{table}
\caption{Prompt template for repairing error in Target Language Code}
\vspace{-0.1in}
\begin{tabularx}{1\textwidth}{X}
\hline
\textbf{\textcolor{purple}{``prompt''}}: 
\textcolor{blue}{\{target\_code\}} \\
Above \textcolor{blue}{\{target\_language\}} has \textcolor{blue}{\{compilation, runtime, infinite loop\}} error. Error Info is given below:\\
\textcolor{blue}{\{err\_context\}} \\
Fix the error and print only the \textcolor{blue}{\{target\_language\}} code and end with the comment $\backslash$``End of Code$\backslash$''. Do not give any other explanation.
\textbf{\textcolor{purple}{``output''}}: 
\textcolor{blue}{\{Answer\}} \\
\hline
\end{tabularx}
\label{tab:repair_prompt}
\end{table}
\noindent We follow a prior work by Pan et al. 2024 to fix errors in translated code. In particular, four common categories of translation errors are identified by Pan et al. 2024, namely compilation errors, test mismatch errors, runtime errors, and infinite loop or hanging input errors.  Our repair strategy is similar to the iterative repair pipeline used by \citep{pan2024lost}, who incorporate multiple iterative rounds of fixes for compilation, runtime, and behavioral errors. The prompt template used for these repairs is provided in \ref{tab:repair_prompt}. The four types of errors are described as follows.

\begin{itemize}
    \item \textbf{Compilation failures.} Often result from missing imports, malformed statements, misplaced brackets, type-checking inconsistencies, or syntactic deviations introduced during translation. Whenever the translated program fails to compile, our system constructs a repair prompt that contains the erroneous target-language code along with the exact compiler diagnostic. The LLM is then instructed to fix only the reported issues and regenerate the corrected code. We employ an iterative repair strategy in which each iteration attempts to eliminate any remaining compilation errors, terminating once the code compiles successfully and passes all the test cases. If the compilation error gets fixed but another error emerges, we iteratively try to fix that error too.
    \item \textbf{Test mismatch errors.} These occur when a translated program compiles successfully but produces incorrect outputs for one or more test cases. Unlike compilation failures, test mismatches expose semantic inaccuracies introduced during translation, such as incorrect branching logic, mishandling of edge cases, data structure inconsistencies, or subtle arithmetic deviations. To repair such errors, the repair prompt includes the failing input, the expected output, and the observed incorrect output produced by the translated code. With this structured feedback, the LLM is encouraged to refine the program logic to match the intended behavior defined by the original problem specification. 
    \item \textbf{Runtime errors.} include division by zero, index-out-of-bounds exceptions, null de-references, and type conversion failures. These errors typically reflect deeper structural flaws in the translated code. When a runtime exception occurs, the repair prompt includes the full stack trace, the line that triggered the error, and the failing input. This contextual information helps the LLM identify the source of the exception and reconstruct the code to ensure safe and correct execution.
    \item \textbf{Infinite-loop or hanging-input errors.} These issues arise when the translated program fails to terminate or waits indefinitely for input, often due to incorrect loop termination conditions, misplaced update operations, or misinterpreted input-handling logic. In such cases, our system signals a timeout and provides the LLM with the partial execution context, including the loop structure and the state at the time of interruption. The model is then prompted to adjust the loop bounds, update statements, or input operations to ensure termination. 

\end{itemize}

\noindent The workflow of this iterative strategy is as follows: after a translation is generated, the system first attempts to compile the code. If a compilation error is detected, the LLM is invoked to repair the issue, with up to three repair attempts allowed. Once compilation succeeds, the translated code is executed against all available test cases. If any test case fails or if the execution triggers a runtime exception or an infinite-loop timeout, the system again initiates up to three iterations of error-specific repair. For test mismatches, the LLM receives the failing input, expected output, and observed output. For runtime or infinite-loop errors, it receives the full diagnostic trace and execution context. In all cases, the iterative loop terminates as soon as the error is successfully resolved or once the maximum of three repair attempts has been reached. This bounded and uniform iteration policy ensures that different categories of errors are treated consistently while giving the LLM sufficient opportunity to correct both syntactic and semantic faults introduced during translation.
}
{

\subsection{Analyzing the Quality of Translated Code}
\label{sonar_run}
We evaluate the quality of translated code by analyzing both fixed and non-fixed versions of \specS and their corresponding source code translations using SonarQube~\citep{sonarqube}. Our goal is to assess translation quality rather than stylistic differences across languages. Accordingly, we rely on SonarQube-reported high-severity defects that are more likely to reflect semantic, security, or correctness-related issues introduced during translation. The analysis spans five programming languages, three datasets, and two translation approaches. SonarQube is a widely used static analysis platform that reports potential issues related to code quality, security, and maintainability, making it suitable for evaluating the structural and semantic properties of automatically translated programs.

\textbf{Analysis Configuration.} All analyses are conducted using SonarQube via SonarCloud (free tier) with \texttt{sonar-scanner-cli-8.0.1.6346}, running on a Linux environment. We use the default \emph{sonar-way} rule set and apply the default quality gate uniformly across all languages and datasets to ensure consistency. SonarQube reports issues across multiple dimensions, including \emph{Security}, \emph{Reliability}, \emph{Maintainability}, \emph{Duplications}, \emph{Size}, and \emph{Complexity}, with each issue assigned one of five severity levels: \emph{Blocker}, \emph{Critical}, \emph{Major}, \emph{Minor}, and \emph{Info}.  
To reduce sensitivity to coding style and superficial conventions, we focus exclusively on \emph{Blocker} and \emph{Critical} issues, which predominantly correspond to defects affecting correctness, security, or runtime behavior. Lower-severity issues, which are more likely to capture stylistic or convention-related concerns, are excluded from the analysis.

\textbf{Normalization and Scope.} Translated programs vary considerably in size across languages and datasets. To enable fair comparisons, we normalize issue counts by reporting the number of \emph{Blocker} and \emph{Critical} issues per 1,000 non-commented lines of code (NCLOC), where NCLOC values are computed by SonarQube. Only successfully analyzable files are included in the analysis; files that fail to compile (when compilation is required) are excluded.

\textbf{Language-Specific Setup.} SonarQube requires different analysis pipelines depending on the programming language. We therefore adopt three distinct setups:

\begin{itemize}
    \item \textbf{Java.} SonarQube performs Java analysis at the bytecode level and thus requires compiled binaries. For each Java project, all \texttt{.java} files are compiled using \texttt{javac} prior to analysis. Only successfully compiled files are analyzed. The SonarQube scanner is then executed with the compiled binaries specified via the \texttt{sonar.java.binaries} parameter.


    \item \textbf{C/C++.} SonarQube requires detailed build information to accurately analyze C and C++ programs, as many defects depend on how code is compiled. We therefore use SonarQube’s \emph{build wrapper}, a utility that intercepts compiler invocations during the build process and records how each source file is compiled. For each project, all translated C/C++ source files are placed under a \texttt{Code/} directory, with each file located in its own subdirectory to enable independent compilation.

    A shared \texttt{CMakeLists.txt} file is used to configure the build. It recursively collects all \texttt{.c}, \texttt{.cpp}, and \texttt{.c++} files under \texttt{Code/} and creates a separate executable target for each source file. The project is configured with \texttt{cmake} and compiled with \texttt{make}, with the build executed through the build wrapper (\texttt{build-wrapper-linux-x86-64}).

    During compilation, the build wrapper records, for each source file, a \emph{compile command}, which specifies the compiler binary (e.g., \texttt{gcc} or \texttt{clang}), the compilation flags, the \emph{include paths} used to resolve header files, and any macro definitions. These compile commands are stored in a standard \texttt{compile\_commands.json} file and provided to SonarQube via the \texttt{sonar.cfamily.compile-commands} parameter. This information allows SonarQube to analyze each source file under the same compilation context as in the actual build, thereby improving analysis precision and reproducibility.

    \item \textbf{Python and Go.} Python and Go do not require a pre-compilation step for SonarQube analysis. As long as the source files are syntactically valid, the SonarQube scanner can analyze them directly. The analysis operates at the source level and reports issues without requiring additional build artifacts.
\end{itemize}

\textbf{Cloud-Based Reporting and Data Collection.} For each project, analysis results are uploaded to SonarCloud using the SonarQube scanner. We then retrieve detailed issue statistics, severity distributions, and NCLOC measurements programmatically via the SonarCloud Web API. This automated API-based data collection ensures consistency, minimizes manual intervention, and enables scalable aggregation of results across languages, datasets, and translation variants.

The extracted data are aggregated per project and normalized by NCLOC to compute issue densities, which serve as the basis for our comparative evaluation of fixed versus non-fixed translations.
}

{
\subsection{Environment Setup}
\noindent All experiments were conducted on a SLURM-managed computing cluster. For experiments using the local Magicoder model, we allocated two GPUs per run. The available GPU nodes were equipped with NVIDIA A100 and NVIDIA V100 GPUs. Each run processed a single language pair, and multiple language pairs were executed in parallel across different nodes. For each job, we requested 60\,GB of system RAM.

The inference time for each LLM call with the local model ranged from approximately 5 seconds to 1 minute, occasionally longer depending on the GPU type (A100 or V100), node load, and other system-level factors.

For API-based models (GPT-4 and DeepSeek-Coder), the latency per call ranged from approximately 5 seconds to 30 seconds, depending on network conditions and server-side factors beyond our control. Using GPT-4 incurs a cost of \$23.11, while using DeepSeekCoder costs \$4.32.
}

\section{Evaluation Results}
\label{eval_res}
In this section, we provide the evaluation results to answer the three RQs.

\pa{Evaluation Settings.}
{ For our experiments, we evaluate three LLMs: GPT-4~\citep{achiam2023gpt}, DeepSeek-Coder~\citep{zhu2024deepseek}, and Magicoder-6.7B~\citep{wei2023magicoder}, described in Section~\ref{llms}, across three datasets for code translation. Since GPT-4 achieved the best performance in prior work~\citep{pan2024lost}, we select it for a fair comparison with the baseline. DeepSeek-Coder and Magicoder are additionally included to assess the reproducibility and generalizability of our approach.} To generate the responses, we used a temperature of 0.7 as it provides a good balance between creativity and coherence. A lower temperature would result in more deterministic and predictable responses, which could be overly rigid for code translation, where some flexibility and adaptation to context are needed. On the other hand, a higher temperature could lead to responses that are too diverse or off-topic, potentially introducing inaccuracies.
{To verify our choice, we conducted a temperature ablation study on GPT-4 with the Avatar-Verified dataset with translation using \specS and source code to decide the best temperature for code translation. We studied three temperature settings: 0.6, 0.7, and 0.8, where we run the LLM using each temperature three times to verify that the observed results are consistent and not due to randomness. Table \ref{tab:temp_ablation} shows the outcome.}

{To evaluate correctness, we use Computational Accuracy (CA), a standard metric commonly used in previous works \citep{pan2024lost, macedo2024intertrans, yuan2024transagentllmbasedmultiagentcode}, to facilitate direct comparison. CA is defined as the ratio of datapoints that pass all corresponding test cases to the total number of datapoints.} We consider a translation passed only if all test cases pass for the translated code.

{We selected C, C++, Go, Java, JavaScript, Rust, and Python as target languages. We used gcc v11.4.0, g++ v11.4.0, OpenJDK 11.0.25, go v1.23.2 Linux/amd64, node v12.22.9, rustc v1.92.0, and python v3.10.13 to evaluate the translations to C, C++, Java, Go, Javascript, Rust, and Python, respectively.} The operating system is Ubuntu 22.04.4 LTS.


\begin{table*}

\centering
\scriptsize
\renewcommand{\arraystretch}{1.1} 
\caption{Code translation results using GPT-4, compared to the LIT \citep{pan2024lost}. Values in parentheses with ($\uparrow$) indicate the absolute improvement in correctness after error fixing, compared to the original translation results without fixing.}
\resizebox{\textwidth}{!}{  
\begin{tabular}{|m{1.3cm}|m{1.2cm}|m{2.8cm}|m{1cm}|m{2.2cm}|m{2.2cm}|m{2.2cm}|}
\hline
\textbf{Dataset} & \textbf{Source Language} & \textbf{Target Language(s)} & \textbf{Total Codes} & \textbf{LIT} & \textbf{With Only \specS} & \textbf{With \specS+ Source} \\
\hline

\multirow{2}{*}{Avatar-} & Java & C, C++, Go, Python & 960 & 78.75\% (4.7\%$\uparrow$) & 73.125\% (28.11\%$\uparrow$) & 83.854\% (16.51\%$\uparrow$) \\
\cline{2-7}
 Verified & Python & C, C++, Go, Java & 956 & 56.9\% (7.1\%$\uparrow$) & 61.514\% (21.52\%$\uparrow$) & 65.167\% (14.47\%$\uparrow$) \\
\hline
\multirow{5}{*}{CodeNet} & C & C++, Go, Java, Python & \multirow{5}{*}{4000} & 88.625\% (6.78\%$\uparrow$) & 70.75\% (14.11\%$\uparrow$) & 84.625\% (12.65\%$\uparrow$)\\
\cline{2-3} \cline{5-7}
 & C++ & C, Go, Java, Python & & 84.625\% (5.78\%$\uparrow$) & 77.5\% (12.12\%$\uparrow$) & 87.625\% (9.74\%$\uparrow$)\\
 \cline{2-3} \cline{5-7}
 & Go & C, C++, Java, Python & & 91.125\% (6.58\%$\uparrow$) & 77.625\% (16.07\%$\uparrow$) & 86.125\% (8.68\%$\uparrow$) \\
 \cline{2-3} \cline{5-7}
 & Java & C, C++, Go, Python & & 84.375\% (3.78\%$\uparrow$) & 76.375\% (25.72\%$\uparrow$) & 83.25\% (12.5\%$\uparrow$) \\
 \cline{2-3} \cline{5-7}
 & Python & C, C++, Go, Java & & 83.625\% (4.66\%$\uparrow$) & 83.125\% (7.34\%$\uparrow$) & 89.5\% (7.83\%$\uparrow$) \\
\hline
Evalplus & Python & Java & 164 & 84.76\% (10.32\%$\uparrow$) & 79.268\% (2.36\%$\uparrow$) & 85.37\% (6.87\%$\uparrow$) \\
\hline \hline
Total/ Average & - & - & 6080 & 81.6\% (6.17\%$\uparrow$) & 74.91\% (15.6\%$\uparrow$) & 83.19\% (10.7\%$\uparrow$) \\
\hline
\end{tabular}
}
\label{tab:error_repair_results_gpt4}
\end{table*}

\begin{table*}

\centering
\scriptsize
\renewcommand{\arraystretch}{1.1} 
\caption{Code translation using DeepSeek-Coder. Values in parentheses with ($\uparrow$) indicate the absolute improvement in correctness after error fixing, compared to the original translation results without fixing.}
\resizebox{\textwidth}{!}{  
\begin{tabular}{|m{1.2cm}|m{1.2cm}|m{2cm}|m{1cm}|m{1.4cm}|m{1.85cm}|m{1.85cm}|}
\hline
\textbf{Dataset} & \textbf{Source Language} & \textbf{Target Language(s)} & \textbf{Total Datapoints}& \textbf{With Only Source} & \textbf{With Only \specS} & \textbf{With \specS+ Source} \\
\hline

\multirow{2}{*}{Avatar-} & Java & C, C++, Go, Python & 960 & 97.08\% (4.16\%$\uparrow$) & 92.5\% (29.58\%$\uparrow$) & 96.25\% (4.17\%$\uparrow$)\\
\cline{2-7}
 Verified & Python & C, C++, Go, Java & 956 & 97.92\% (14.59\%$\uparrow$) & 98.33\% (16.25\%$\uparrow$) & 98.75\% (13.33\%$\uparrow$)\\
\hline

 \multirow{5}{*}{CodeNet} & C & C++, Go, Java, Python & \multirow{5}{*}{4000} & 97.12\% (5.74\%$\uparrow$) & 95.62\% (10.12\%$\uparrow$) & 97.0\% (5.38\%$\uparrow$)\\
\cline{2-3} \cline{5-7}
 & C++ & C, Go, Java, Python &  & 98.62\% (8.5\%$\uparrow$) & 95.25\% (7.63\%$\uparrow$) & 97.0\% (5.0\%$\uparrow$)\\
 \cline{2-3} \cline{5-7}
 & Go & C, C++, Java, Python &  & 97.25\% (7.37\%$\uparrow$) & 94.5\% (10.0\%$\uparrow$) & 96.38\% (6.63\%$\uparrow$)\\
 \cline{2-3} \cline{5-7}
 & Java & C, C++, Go, Python &  & 96.12\% (4.87\%$\uparrow$) & 93.88\% (12.5\%$\uparrow$) & 95.0\% (6.62\%$\uparrow$)\\
 \cline{2-3} \cline{5-7}
 & Python & C, C++, Go, Java &  &  98.5\% (2.12\%$\uparrow$) & 97.5\% (5.88\%$\uparrow$) & 98.75\% (2.25\%$\uparrow$)\\
\hline
Evalplus & Python & Java & 164 & 81.1\% (29.27\%$\uparrow$) & 82.32\% (36.59\%$\uparrow$) & 82.93\% (31.71\%$\uparrow$)\\
\hline \hline
Total/   Average & -- & -- & 6080 & 97.1\% & 95.3\% & 96.7\% \\
\hline 
\end{tabular}
}
\label{tab:deepseek_res}
\end{table*}

\begin{table*}

\centering
\footnotesize
\setlength{\tabcolsep}{3pt}
\renewcommand{\arraystretch}{1.15}
\caption{ Code translation result on CodeNet (Dataset sample from Intertrans) using MagiCoder, compared with InterTrans~\citep{macedo2024intertrans}.Values in parentheses ($\uparrow$) denote absolute correctness improvement after error fixing. Results are reported over all target languages and excluding Rust and JavaScript. `All' columns represent the results across all target languages, while `w/o' column excludes JavaScript and Rust from the target languages.}

\resizebox{\textwidth}{!}{
\setlength{\extrarowheight}{4pt}  
\begin{tabular}{|
>{\RaggedRight\arraybackslash}p{1.8cm}|
>{\centering\arraybackslash}p{1.5cm}|
>{\RaggedRight\arraybackslash}p{3.4cm}|
>{\centering\arraybackslash}p{1.0cm}|
>{\centering\arraybackslash}p{1.25cm}|
>{\centering\arraybackslash}p{1.25cm}|
>{\centering\arraybackslash}p{1.25cm}|
>{\centering\arraybackslash}p{1.25cm}|
>{\centering\arraybackslash}p{1.25cm}|
>{\centering\arraybackslash}p{1.25cm}|
>{\centering\arraybackslash}p{1.25cm}|
>{\centering\arraybackslash}p{1.25cm}|}
\hline
\textbf{Dataset} & \textbf{Source} & \textbf{Targets} & \textbf{N} &
\multicolumn{2}{c|}{\parbox[c]{2.5cm}{\centering\textbf{Only Source}}} &
\multicolumn{2}{c|}{\parbox[c]{2.5cm}{\centering\textbf{Only \specS}}} &
\multicolumn{2}{c|}{\parbox[c]{2.5cm}{\centering\textbf{\specS + Source}}} &
\multicolumn{2}{c|}{\parbox[c]{2.5cm}{\centering\textbf{InterTrans}}} \\
\cline{5-12}
 & & & &
\textbf{All} & \textbf{w/o} &
\textbf{All} & \textbf{w/o} &
\textbf{All} & \textbf{w/o} &
\textbf{All} & \textbf{w/o} \\
\hline

\multirow{6}{*}{\parbox{1.6cm}{\centering CodeNet\\(InterTrans)}} 
& C++
& \parbox{3cm}{Go, Java, Python, JavaScript, Rust}
& \multirow{6}{*}{1050}
& \parbox{1.25cm}{57.71\\(10.28$\uparrow$)} & 82.86
& \parbox{1.25cm}{34.29\\(5.15$\uparrow$)}  & 51.43
& \parbox{1.25cm}{54.29\\(8.00$\uparrow$)}  & 80.00
& 88.00 & 94.29 \\
\cline{2-3}\cline{5-12}

& Go
& \parbox{3cm}{C++, Java, Python, JavaScript, Rust}
&
& \parbox{1.25cm}{59.29\\(8.58$\uparrow$)}  & 74.29
& \parbox{1.25cm}{42.86\\(7.15$\uparrow$)}  & 50.48
& \parbox{1.25cm}{69.29\\(12.86$\uparrow$)} & 80.95
& 85.71 & 90.48 \\
\cline{2-3}\cline{5-12}

& Java
& \parbox{3cm}{C++, Go, Python,\\ JavaScript, Rust}
&
& \parbox{1.25cm}{55.43\\(18.86$\uparrow$)} & 73.33
& \parbox{1.25cm}{30.29\\(9.15$\uparrow$)}  & 41.90
& \parbox{1.25cm}{49.14\\(13.71$\uparrow$)} & 67.62
& 85.14 & 91.43 \\
\cline{2-3}\cline{5-12}

& Python
& \parbox{3cm}{C++, Go, Java,\\ JavaScript, Rust}
&
& \parbox{1.25cm}{54.29\\(18.29$\uparrow$)} & 75.24
& \parbox{1.25cm}{32.57\\(6.28$\uparrow$)}  & 49.52
& \parbox{1.25cm}{52.57\\(15.43$\uparrow$)} & 76.19
& 91.43 & 93.34 \\
\cline{2-3}\cline{5-12}

& JavaScript
& \parbox{3cm}{C++, Go, Java,\\ Python, Rust}
&
& \parbox{1.25cm}{55.43\\(39.28$\uparrow$)} & 62.86
& \parbox{1.25cm}{29.29\\(23.58$\uparrow$)} & 32.38
& \parbox{1.25cm}{57.71\\(44.28$\uparrow$)} & 65.71
& 87.43 & 87.86 \\
\cline{2-3}\cline{5-12}

& Rust
& \parbox{3cm}{C++, Go, Java,\\ Python, JavaScript}
&
& \parbox{1.25cm}{70.48\\(4.77$\uparrow$)}  & 70.48
& \parbox{1.25cm}{40.95\\(5.71$\uparrow$)}  & 40.95
& \parbox{1.25cm}{71.43\\(6.67$\uparrow$)}  & 71.43
& 86.29 & 88.57 \\
\hline
\hline
Total/ Average & -- & -- &
\parbox{1.0cm}{1050} &
58.77 & 72.53 &
35.04 & 43.67 &
59.07 & 73.14 &
87.33 & 90.72 \\
\hline 
\end{tabular}
}
\label{tab:intertrans_full}
\end{table*}

\subsection*{\textbf{RQ1-\RQOne}}
\pa{Motivation} Previous studies have shown that LLMs can translate code across programming languages with reasonable accuracy. However, code translation remains challenging when the source code is complex, long, or highly language-specific. Natural language (NL) specifications offer a higher-level description of program intent and may help LLMs focus on semantics rather than syntax. If LLM-generated specifications can capture the essential behavior of a program, they may serve as useful intermediate representations that improve translation quality. Therefore, this research question investigates whether using LLM-generated NL specifications, either alone or combined with source code can improve LLM-based code translation compared to using source code alone.

\pa{Method} To answer this RQ, we first prompted LLMs to generate the \specS description for a given code from the dataset. Then we used our three strategies to compare the translation capability: \textbf{(1) Prompting LLMs to translate code to a target language using only the \spec (2) Prompting LLMs to translate code to a target language using only the source code and (3) Prompting LLMs to translate code to a target language using a combination of source code and the corresponding \spec.} We compare the results of the three above-mentioned approaches with baselines \citep{pan2024lost} and \citep{macedo2024intertrans}. We have a total of 1916, 4000, and 164 translations from Avatar-Verified \citep{ahmad-etal-2021-avatar}, CodeNet \citep{puri2021codenet}, and Evalplus \citep{liu2024your}, respectively.

{
\noindent \textbf{Results.} Tables~\ref{tab:error_repair_results_gpt4}, \ref{tab:deepseek_res}, and \ref{tab:intertrans_full} report the evaluation results for our three prompting strategies: (i) the source-only baseline~\citep{pan2024lost}, (ii) NL-specification only (\specS), and (iii) joint conditioning on source code and NL-specification (\specS + Source), using GPT-4, DeepSeek-Coder, and Magicoder, respectively. For each model, we include the corresponding published baselines: Pan et al.~\citep{pan2024lost} for GPT-4, InterTrans~\citep{macedo2024intertrans} for Magicoder, and we report DeepSeek-Coder results as a new model without an established prior baseline.

\noindent \textbf{GPT-4.} From Table~\ref{tab:error_repair_results_gpt4}, we observe that after applying our error-repair pipeline, for Python-to-other-language translations across all three datasets: Avatar-Verified, CodeNet, and EvalPlus, the \specS + Source strategy outperforms the baseline (by 3.79\% on Avatar-Verified, 3.05\% on CodeNet, and 3.05\% on EvalPlus), indicating that \specS combined with source code improves Python-to-other-language translation. An additional improvement is observed for C++, where it outperforms the baseline by 1.625\%.

However, for other language pairs, our strategies do not surpass the baseline. Thus, when aggregated across all datasets and language pairs, the source-only baseline remains the strongest overall, suggesting that while \specS can be beneficial in certain scenarios, it does not consistently outperform direct source-based translation.

\noindent \textbf{DeepSeek-Coder.} As shown in Table~\ref{tab:deepseek_res}, DeepSeek-Coder achieves the best overall performance on the Avatar-Verified and CodeNet datasets among the three models, indicating strong native capabilities for cross-language code translation. However, its performance on EvalPlus is comparatively weaker, suggesting that the model may be less robust to stricter or more diverse test suites that emphasize edge cases.

Across all datasets, the performance differences among the three prompting strategies are relatively small. This indicates that DeepSeek-Coder is less sensitive to NL-spec conditioning and is able to recover much of the required semantic information directly from the source code. One possible explanation is that DeepSeek-Coder has been extensively trained on large-scale, multilingual code corpora, enabling it to internalize common programming patterns and semantics without relying heavily on auxiliary natural language guidance.

We observe that combining source code with \specS occasionally provides modest gains for specific language pairs (e.g., Python-to-Java on EvalPlus), suggesting that NL-specification may still offer complementary signals in scenarios involving ambiguous logic or implicit assumptions. Future improvements could focus on selectively applying \specS only when source code complexity exceeds a certain threshold (e.g., deep control flow or sparse documentation), or on improving the faithfulness of \specS generation to better align with DeepSeek-Coder’s internal representations.

\noindent \textbf{Magicoder.} Table~\ref{tab:intertrans_full} on CodeNet (Intertrans) shows that both \specS and \specS + Source underperform the InterTrans baseline~\citep{macedo2024intertrans}. This gap is expected as InterTrans relies on a multi-hop translation paradigm that chains multiple intermediate languages to exploit asymmetric pairwise strengths of the underlying model. A single source-to-target translation may involve up to 313 LLM calls. This strategy is particularly effective for difficult direct pairs, for example, while Magicoder struggles with direct C++$\rightarrow$Rust translation, it performs substantially better on intermediate pairs such as C++$\rightarrow$Java or Python$\rightarrow$JavaScript. By composing these stronger intermediate translations, InterTrans achieves higher end-to-end accuracy, albeit at a very high computational cost.

In contrast, our approach is intentionally constrained to \emph{direct}\\ source$\rightarrow$target translation and evaluated under three controlled settings: source-only, \specS-only, and joint source + \specS conditioning. Each setting usesLLM calls per translation,covering pseudocode generation, translation, and up to three repair steps. This design explicitly isolates the representational and guiding capacity of \specS, rather than maximizing accuracy through auxiliary translation paths or extensive model invocations. As a result, while InterTrans outperforms our method on language pairs that strongly benefit from intermediate hops, the performance gap narrows on pairs where direct translation is already strong. It further supports this observation, as removing Rust and JavaScript from target languages leads to improved performance for our approach, underscoring that InterTrans’s advantage primarily stems from its ability to explore a large space of intermediate language permutations. 

\noindent \textbf{\specS-only vs.\ source-only.} Across all three models, using \specS alone consistently degrades performance relative to the source-only baseline. This suggests that translating from code to NL-specification introduces information loss and potential misrepresentation, particularly for low-level operations, API usage, and language-specific constructs. As a result, \specS-only translation rarely (on Evalplus, by Magicoder from Python to Java) surpasses source-based translation and should be viewed as a complementary signal rather than a replacement for source code.

\noindent \textbf{Language-specific effects.} Another consistent observation across Tables~\ref{tab:deepseek_res}--\ref{tab:intertrans_full} is the performance drop when translating from C or Java to Python or Go. C and Java programs tend to be larger and contain more explicit, language-specific syntax~\citep{puri2021codenet,zheng2023codegeex}. During NL-specification generation, such constructs are more prone to abstraction or omission, which can lead to incorrect or incomplete target code. Improving NL-spec generation for languages with strong syntactic and semantic constraints is left as future work.

\noindent Overall, { our results show that the benefit of \specS is limited to specific models and language pairs; the source-only baseline remains the strongest overall strategy when averaged across all datasets. These findings highlight a trade-off between architectural leverage and interpretability: while multi-hop methods such as InterTrans achieve higher accuracy by exploiting intermediate representations, our approach emphasizes semantic transparency and a controlled analysis of when and why \specS helps. Future work on improving specification fidelity, selective application of \specS, and hybrid integration with multi-hop translation is likely to further close this performance gap.}

\noindent \textbf{Impact of compilation errors and repair.}
Across all three models (GPT-4, DeepSeek-Coder, and Magicoder), we observe four primary categories of errors in the translated programs: (i) compilation errors, (ii) test case mismatches, (iii) runtime errors, and (iv) infinite loops or programs waiting for input. Among these, compilation errors are consistently the most frequent failure mode, regardless of the prompting strategy or model.

To mitigate these issues, we apply an automated post-translation repair step using the prompt shown in Table~\ref{tab:repair_prompt}. As shown in Tables~\ref{tab:error_repair_results_gpt4}, \ref{tab:deepseek_res}, and \ref{tab:intertrans_full}, this repair step leads to measurable improvements in correctness across all models, with particularly noticeable gains for \specS-only and joint \specS + source translations. For example, GPT-4 benefits from error repair with an 8.5\% improvement for \specS-only translations and a 6.1\% improvement for joint conditioning. Similar trends are observed for DeepSeek-Coder and Magicoder, where a substantial fraction of translation failures can be resolved through syntactic and minor structural fixes.

These results suggest that many translation errors stem from surface-level syntactic issues rather than fundamentally incorrect semantic understanding. Moreover, reliance on NL-specification appears to increase susceptibility to such errors, making post-translation repair especially beneficial when \specS is used either alone or in combination with source code. Overall, automated repair serves as an effective and lightweight mechanism to recover correctness without altering the core translation strategy.
}

\begin{table*}
\centering
\scriptsize
\renewcommand{\arraystretch}{1.1}
\caption{
Paired significance, effect size, and complementarity analysis across datasets and language pairs on DeepSeekCoder. SC = source-only translation, NL = translation using \specS, SL = source language, TL = target languages. Fixes ($c$) = problems solved by the second method but not by the first; regressions ($b$) = problems solved by the first method but not by the second. $p$-values are from McNemar's exact test on paired per-problem outcomes. All $p$-values are Holm-Bonferroni adjusted within each dataset family at $alpha=0.05$; cells shaded in green indicate Holm-significant improvements over SC ($c>b$), and red indicates Holm-significant degradations ($c<b$). c/b columns also report Cliff's $delta$ in parentheses; $|delta|<0.147$ negligible, $[0.147,0.33)$ small, $[0.33,0.474)$ medium, $\geq0.474$ large. Logical-OR ensemble columns count a problem as solved if any constituent strategy solves it.
}
\resizebox{\textwidth}{!}{
\begin{tabular}{|m{1.0cm}|m{1.8cm}|m{0.8cm}|m{0.9cm}|m{1.8cm}|m{0.9cm}|m{1.9cm}|m{1.0cm}|m{1.8cm}|m{1.2cm}|m{1.8cm}|}
\hline
\textbf{Dataset} & \textbf{SL$\rightarrow$TL} & \textbf{SC (\%)} & \textbf{p (SC vs NL)} & \textbf{c/b ($\delta$)} & \textbf{p (SC vs NL + SC)} & \textbf{c/b ($\delta$)} & \textbf{p (SC vs NL $\cup$ (NL + SC))} & \textbf{c/b ($\delta$)} & \textbf{p (SC vs SC $\cup$ NL $\cup$ (NL + SC))} & \textbf{c/b ($\delta$)} \\
\hline
\multirow{8}{*}{\parbox{1.0cm}{\centering Avatar\\-\\Verified}} & Java$\rightarrow$C & 96.7 & 0.05 & 4/13 (-0.04) & 0.06 & 5/0 (+0.02) & 0.06 & 5/0 (+0.02) & 0.06 & 5/0 (+0.02) \\
 & Java$\rightarrow$C++ & 98.8 & 1.00 & 1/2 (-0.00) & 1.00 & 1/2 (-0.00) & 1.00 & 1/0 (+0.00) & 1.00 & 1/0 (+0.00) \\
 & Java$\rightarrow$Go & 92.9 & 0.82 & 11/9 (+0.01) & 0.09 & 10/3 (+0.03) & $<0.01$ & 12/1 (+0.05) & \cellcolor{green!40}{$<0.01$} & 12/0 (+0.05) \\
 & Java$\rightarrow$Python & 97.1 & 0.03 & 5/16 (-0.05) & 0.75 & 4/6 (-0.01) & 0.45 & 5/2 (+0.01) & 0.06 & 5/0 (+0.02) \\
 & Python$\rightarrow$C & 89.5 & 0.81 & 8/10 (-0.01) & 0.82 & 11/9 (+0.01) & 0.04 & 12/3 (+0.04) & \cellcolor{green!40}{$<0.01$} & 12/0 (+0.05) \\
 & Python$\rightarrow$C++ & 95.8 & 0.23 & 3/8 (-0.02) & 0.38 & 1/4 (-0.01) & 1.00 & 3/3 (+0.00) & 0.25 & 3/0 (+0.01) \\
 & Python$\rightarrow$Go & 88.3 & 0.10 & 13/24 (-0.05) & 0.86 & 16/14 (+0.01) & 0.02 & 18/6 (+0.05) & \cellcolor{green!40}{$<0.01$} & 18/0 (+0.08) \\
 & Python$\rightarrow$Java & 97.9 & 1.00 & 4/3 (+0.00) & 0.69 & 4/2 (+0.01) & 0.12 & 4/0 (+0.02) & 0.12 & 4/0 (+0.02) \\
\hline
\multirow{19}{*}{Codenet} & C$\rightarrow$C++ & 97.5 & 0.62 & 3/1 (+0.01) & 0.12 & 4/0 (+0.02) & 0.12 & 4/0 (+0.02) & 0.12 & 4/0 (+0.02) \\
 & C$\rightarrow$Go & 95.5 & 0.51 & 6/3 (+0.01) & 0.03 & 6/0 (+0.03) & 0.02 & 7/0 (+0.04) & 0.02 & 7/0 (+0.04) \\
 & C$\rightarrow$Java & 98.5 & 0.04 & 1/8 (-0.04) & 1.00 & 2/1 (+0.01) & 1.00 & 2/1 (+0.01) & 0.50 & 2/0 (+0.01) \\
 & C$\rightarrow$Python & 97.0 & 0.01 & 2/12 (-0.05) & $<0.01$ & 1/13 (-0.06) & 1.00 & 3/3 (+0.00) & 0.25 & 3/0 (+0.01) \\
 & C++$\rightarrow$C & 99.5 & 0.22 & 1/5 (-0.02) & 0.22 & 1/5 (-0.02) & 0.62 & 1/3 (-0.01) & 1.00 & 1/0 (+0.01) \\
 & C++$\rightarrow$Java & 99.5 & 0.07 & 1/7 (-0.03) & 1.00 & 1/0 (+0.01) & 1.00 & 1/0 (+0.01) & 1.00 & 1/0 (+0.01) \\
 & C++$\rightarrow$Python & 95.5 & 0.03 & 5/16 (-0.06) & 0.12 & 4/11 (-0.04) & 0.34 & 7/3 (+0.02) & 0.02 & 7/0 (+0.04) \\
 & Go$\rightarrow$C & 96.5 & 0.23 & 3/8 (-0.03) & 1.00 & 5/4 (+0.01) & 0.22 & 5/1 (+0.02) & 0.06 & 5/0 (+0.03) \\
 & Go$\rightarrow$C++ & 98.5 & 0.29 & 2/6 (-0.02) & 1.00 & 2/3 (-0.01) & 1.00 & 2/1 (+0.01) & 0.50 & 2/0 (+0.01) \\
 & Go$\rightarrow$Java & 100.0 & 0.06 & 0/5 (-0.03) & 0.25 & 0/3 (-0.01) & 1.00 & 0/1 (-0.01) & 1.00 & 0/0 (+0.00) \\
 & Go$\rightarrow$Python & 94.0 & 0.08 & 4/12 (-0.04) & 0.45 & 6/10 (-0.02) & 1.00 & 6/5 (+0.01) & 0.03 & 6/0 (+0.03) \\
 & Java$\rightarrow$C & 97.5 & 1.00 & 4/4 (+0.00) & 0.62 & 3/1 (+0.01) & 0.12 & 4/0 (+0.02) & 0.12 & 4/0 (+0.02) \\
 & Java$\rightarrow$C++ & 98.0 & 0.06 & 0/5 (-0.03) & 0.62 & 1/3 (-0.01) & 1.00 & 1/1 (+0.00) & 1.00 & 1/0 (+0.01) \\
 & Java$\rightarrow$Go & 93.5 & 1.00 & 6/6 (+0.00) & 1.00 & 6/6 (+0.00) & 0.07 & 9/2 (+0.04) & $<0.01$ & 9/0 (+0.04) \\
 & Java$\rightarrow$Python & 95.5 & $<0.01$ & 3/16 (-0.07) & 0.02 & 2/11 (-0.04) & 0.51 & 3/6 (-0.01) & 0.25 & 3/0 (+0.01) \\
 & Python$\rightarrow$C & 98.0 & 1.00 & 3/3 (+0.00) & 1.00 & 2/1 (+0.01) & 0.62 & 3/1 (+0.01) & 0.25 & 3/0 (+0.01) \\
 & Python$\rightarrow$C++ & 99.5 & 0.50 & 0/2 (-0.01) & 1.00 & 0/0 (+0.00) & 1.00 & 0/0 (+0.00) & 1.00 & 0/0 (+0.00) \\
 & Python$\rightarrow$Go & 97.5 & 0.11 & 2/8 (-0.03) & 1.00 & 1/1 (+0.00) & 0.50 & 2/0 (+0.01) & 0.50 & 2/0 (+0.01) \\
 & Python$\rightarrow$Java & 99.0 & 1.00 & 1/1 (+0.00) & 1.00 & 1/0 (+0.01) & 1.00 & 1/0 (+0.01) & 1.00 & 1/0 (+0.01) \\
\hline
\multirow{1}{*}{Evalplus} & Python$\rightarrow$Java & 81.1 & 0.73 & 5/3 (+0.01) & 0.45 & 5/2 (+0.02) & 0.07 & 7/1 (+0.04) & 0.02 & 7/0 (+0.04) \\
\hline
\end{tabular}
}
\label{tab:significance_result}
\end{table*}
\begin{table*}

\centering
\scriptsize
\renewcommand{\arraystretch}{1.2} 
\caption{Temperature ablation study using GPT-4}
\resizebox{\textwidth}{!}{  
\begin{tabular}{|m{1.3cm}|m{1.2cm}|m{2.8cm}|m{1cm}|m{2.2cm}|m{2.2cm}|m{2.2cm}|}
\hline
\multirow{2}{*}{\textbf{Dataset}} & \multirow{2}{*}{\textbf{Source}} & \multirow{2}{*}{\textbf{Target}} & \multirow{2}{*}{\textbf{Total}} & \multicolumn{3}{|c|}{\textbf{\specS+Source Correctness}} \\ \cline{5-7}
 & & & & temp = 0.6 & temp = 0.7 & temp = 0.8\\
\hline
\multirow{2}{*}{\parbox{1.2cm}{\centering Avatar-\\Verified}} & Java & C, C++, Go, Python & 960 & 65.52\% & 71.97\% & 70.1\% \\ \cline{2-7}
 & Python & C, C++, Go, Java & 956 & 52.62\% & 56.93\% & 55.13\% \\
\hline
\end{tabular}
}
\label{tab:temp_ablation}
\end{table*}

\begin{boxFindings}
    Finding 1.1: Utilizing \specS as the sole intermediate representation for code translation does not outperform direct code-to-code translation, suggesting that LLMs require the structural context of the source code to maintain logical fidelity.
\end{boxFindings}

\begin{boxFindings}
    {Finding 1.2: Combining \specS with source code does not yield a general improvement. Gains are language-pair--specific and source-language--dependent (concentrated on Python and C++ sources), while accuracy degrades for other source languages such as C and Go. The benefit of \specS should be treated as complementary to the source-only approach, not universal.}
\end{boxFindings}
\noindent
\begin{figure*}
\centering
\tiny
    \begin{tabular}{|p{\linewidth}|}
    \hline
        \textbf{\footnotesize Source Code $(HumanEval\_108.py)$} \\ \hline
        \begin{lstlisting}[escapechar=!]
......
1. if l[0] == "-":
2.     l = l[1:]
3.     l = !\colorbox{targetMark}{list(map(int, l))}!
4.     !\colorbox{targetMark}{l[0] = -l[0]}!
5. else:
6.     l = !\colorbox{targetMark}{list(map(int, l))}!
7. return 1 if sum(l) > 0 else 0
...... \end{lstlisting} \\ \hline        
        \textbf{\footnotesize Generated \spec} \\ \hline
...... \\
\pseudocode{\tiny 1.  IF first character of l is "-" THEN:}\\
\pseudocode{\tiny 2.  \space REMOVE first character from l}\\
\pseudocode{\tiny 3.  \space CONVERT} \colorbox{targetMark}{\pseudocode{\tiny remaining characters of l to integers and store in a list}} \\
\pseudocode{\tiny 4.   \space \colorbox{targetMark}{SET first element of the list to its negative value}}\\
\pseudocode{\tiny 5.  ELSE:}\\
\pseudocode{\tiny 6.  \space CONVERT characters of l to integers and store in a list}\\
\pseudocode{\tiny 7.  RETURN 1 IF sum of elements in the list > 0 ELSE return 0}\\
...... \\ \hline
        
        \textbf{\footnotesize Translation with Source Only (Incorrect Translation)} \\ \hline
        \begin{lstlisting}[escapechar=~]
......
1 . if (x < 0) {
2 .     x = -x;
3 .     int firstDigit = -Integer.parseInt(Integer.toString(x).substring(0, 1));
4 .     int otherDigits = ~\colorbox{wrongMark}{Integer.parseInt(Integer.toString(x).substring(1));}~
5 .     if ~\colorbox{wrongMark}{(firstDigit + otherDigits > 0)}~ count++;
6 . } else {
7 .     int sum = 0;
8 .     while (x != 0) {
9 .         sum += x % 10;
10.         x = x / 10;
11.     }
12.     if (sum > 0) count++;
14. }
......
        \end{lstlisting} \\ \hline        
        \textbf{\footnotesize Translation with \spec Only (Correct Translation)} \\ \hline
        \begin{lstlisting}[escapechar=!]
......
1. if (l.charAt(0) == '-') {
2.     l = l.substring(1);
3.     for (char c : l.toCharArray()) {
4.         !\colorbox{correctMark}{list.add(Character.getNumericValue(c));}!
5.     }
6.     !\colorbox{correctMark}{list.set(0, -list.get(0));}! // Set first element to its negative value
7. } 
......
        \end{lstlisting} \\ \hline
    \end{tabular}
    \caption{How correct \specS can boost translation. 
    }
    \label{fig:sample_pseudocode_improvement_example}
\end{figure*}

\noindent
\begin{figure}
\centering
\tiny
    \begin{tabular}{|p{\linewidth}|}
        \hline
        \textbf{\footnotesize Source Code $(HumanEval\_63.py)$} \\ \hline
        \begin{lstlisting}[escapechar=!]
......
1. for _ in range(3, n + 1):
2.     !\colorbox{targetMark}{a, b, c = b, c, a + b + c}!
......
        \end{lstlisting} \\ \hline
        \textbf{Generated \spec} \\ \hline
...... \\
\pseudocode{\tiny 1. FOR i FROM 3 TO n DO} \\
\pseudocode{\tiny 2. \space\space SET temp TO a} \\
\pseudocode{\tiny 3. \space\space SET a TO b} \\
\pseudocode{\tiny 4. \space\space SET b TO c} \\
\pseudocode{\tiny 5.\space\space\colorbox{targetMark}{\tiny SET c TO temp + b + c}}\\
\pseudocode{\tiny 6. ENDFOR} \\
...... \\ \hline
        \textbf{\footnotesize Translation with Source Only (Correct)} \\ \hline
        \begin{lstlisting}[escapechar=!]
......
1. for (int i = 3; i <= n; i++) {
2.     int temp = a;
3.     a = b;
4.     b = c;
5.     !\colorbox{correctMark}{c = temp + a + b;}!
6. }
......
        \end{lstlisting} \\ \hline
        \textbf{\footnotesize Translation with \spec (Incorrect)} \\ \hline
        \begin{lstlisting}[escapechar=!]
......
1. for (int i = 3; i <= n; i++) {
2.     int temp = a;
3.     a = b;
4.     b = c;
5.     !\colorbox{wrongMark}{c = temp + b + c;}!
6. }
...... 
        \end{lstlisting} \\ \hline
    \end{tabular}
    \caption{\footnotesize How incorrect \specs deteriorate translation.}
    \label{fig:sample_pseudocode_deteriorate}
\end{figure}


\subsection*{\textbf{RQ2-\RQTwo}}
{\pa{Motivation.} RQ1 evaluates the quantitative impact of conditioning LLM-based code translation on \specS, but does not explain why \specS helps in some cases and degrades performance in others. RQ2 complements RQ1 by qualitatively analyzing how the content and correctness of \specS influences translation outcomes, using the same LLMs and translation settings. By examining representative cases where \specS improves or downgrades translations relative to source-only prompting, RQ2 aims to uncover concrete success patterns and failure modes that explain the trends observed in RQ1.}

{\pa{Method.} To answer RQ2, we first conducted a manual analysis on 100 translations. For each source language (C, C++, Go, Java, and Python), we selected ten translations where using \specS led to correct results while using source code did not, and ten where the opposite occurred, \specS produced incorrect results while source code succeeded. This resulted in 20 samples per language, covering both cases of \specS impact. 

We also conducted a manual analysis on 250 generated \specS (50 for each programming language from C, C++, Go, Java, Python).} Two annotators, both co-authors of this study and graduate students with over six years of professional software development experience, independently analyzed the selected samples using thematic coding. Thematic coding is a qualitative analysis method in which recurring patterns or themes are iteratively identified, refined, and categorized based on close inspection of the data, rather than being predefined in advance \citep{braun2006using}. Each annotator first performed open coding to identify potential themes related to the effect of \specS on translation outcomes, followed by iterative refinement to consolidate and name the final themes.

Inter-annotator agreement was assessed using Cohen’s kappa, yielding a substantial level of agreement ($\kappa = 0.81$) \citep{cohen1960coefficient,landis1977measurement}. Disagreements were subsequently resolved through structured discussion: annotators jointly reviewed conflicting labels, justified their original decisions with reference to the code and specification behavior, and reached a consensus on the most appropriate theme. The agreed-upon labels were then used in the final analysis.

The 250 sample size is statistically significant at a 95\% confidence level with a 5.7\% margin of error. 

\pa{Results.} { After manually analyzing 250 NL-specification samples, we found that 87.2\% (218/250) were correct, while 12.8\% (32/250) were incorrect. For the 32 incorrect NL-specification instances, the \textit{\specS-only} strategy always results in incorrect translations, whereas combining source code with NL-specification (\textit{source+\specS}) leads to correct translations in 15 cases.

Among the 232 instances with correct NL specifications, 81.0\% (188/232) produced correct translations when using NL-specification only, while 84.1\% (195/232) produced correct translations when combining source code with NL-specification. This shows that when NL-specification is generated correctly, it improves translation accuracy, and that combining NL-specification with source code further increases robustness by mitigating errors introduced by imperfect specifications.

In the separate manual inspection of 100 translated instances, we observed that \textit{\specS-only} produced correct translations in cases where the LLM-generated \specS correctly simplified complex source code, while the corresponding translations generated using \textit{source-code-only} were incorrect. This suggests that simplified explanations of complex source code can enable correct translations.}

Figure~\ref{fig:sample_pseudocode_improvement_example} shows an example from EvalPlus~\citep{liu2024your} (problem HumanEval\_108), where \specs leads to a correct translation. For a negative number (e.g., \textit{-512}), the source code snippet calculates the sum as \textit{(-5)+1+2}. The LLM-generated \specS clearly describes the process in a simple way by stating that the first digit of a negative number should be negated (lines 3-4). 
When translating using only the source code, GPT-4 fails to understand the original complex logic and approaches different summation logic for negative numbers. 
In the same example, for \textit{-512}, the translated code summarizes as \textit{(-5)+12}. What is worse is that the translation using source code only is prone to runtime errors as it tries to make a substring from index 1 (line 4), even for a string of length 1.

For the cases where translation using \specS was incorrect but the translation using source code was correct, we found that the generated \specS was not correct and the error in \specS mainly happens because of the logical reasoning limitations of LLM.

We present such a case in Figure~\ref{fig:sample_pseudocode_deteriorate} to illustrate how an incorrect intermediate \spec can deteriorate translation quality. In the source code, Python performs a \emph{simultaneous tuple assignment}, where the variables $a$, $b$, and $c$ are reassigned to $b$, $c$, and $a+b+c$, respectively (line~2). This operation evaluates the right-hand side expressions using the original values of all variables before any reassignment occurs.

In the generated \specS, this tuple assignment is decomposed into a sequence of imperative steps (lines~2--5). While lines~2--4 correctly introduce a temporary variable and update $a$ and $b$, line~5 incorrectly assigns $c$ the value \texttt{temp + b + c}. At this point, both $b$ and $c$ have already been updated, causing the computation to deviate from the original semantics of $a+b+c$. This error arises from failing to preserve the simultaneous-assignment semantics of Python when linearizing the logic into a generic \specS.

During translation, the LLM correctly follows the logic specified in the generated \specS and produces translated code that faithfully mirrors lines~2--5 of the \specS. However, because the \specS itself is semantically incorrect, the resulting translated code is also incorrect. This example demonstrates that once an error is introduced at the \specS generation stage, it inevitably propagates to the final translation. Consequently, even a faithful translation process cannot recover correctness when relying on a flawed \spec.

{
\noindent \textbf{Summary of success and failure regimes.}
Across our manual analysis, two clear regimes emerge.
\textit{NL-spec helps} when the source code contains complex logic (e.g., non-trivial control flow, simultaneous assignments) that the LLM struggles to translate directly, and when the generated \specS correctly simplifies that logic into an explicit, step-by-step form.
\textit{NL-spec hurts} when the \specS itself is incorrect, errors introduced at the specification stage propagate deterministically into the translation, and even a faithful translation process cannot recover correctness from a flawed \spec.
The dominant factor is therefore not whether \specS is used, but \emph{whether it is generated correctly}.
}

\begin{boxFindings}
    {Finding 2: The effectiveness of \specS is conditional on the correctness of the generated specification. When \specS correctly captures program semantics, it assists translation, particularly for structurally complex programs. When \specS is incorrect, errors propagate deterministically into the translation, making it worse than source-only. \specS should therefore be treated as a complementary signal whose benefit depends on specification quality, not as a universally reliable intermediate representation.}
\end{boxFindings}

\noindent \textbf{Pairwise significance and complementarity analysis across datasets.}
To ensure the statistical reliability of our findings and to distinguish genuine improvements from random variance, we conduct pairwise significance testing using McNemar’s exact test on per-problem outcomes.
We focus this analysis on DeepSeekCoder, which demonstrates the strongest overall performance and thus provides the most rigorous baseline for comparison.

Because high baseline accuracy can limit the statistical power of aggregate metrics, we further analyze the distribution of \emph{fixes} ($c$) versus \emph{regressions} ($b$).
When the source-only strategy already performs very well, only a small number of problems remain unsolved.
In such cases, even if NL-based strategies fix some of these remaining errors, the total number of improvements may be too small for the $p$-value to reach statistical significance.
For this reason, non-significant $p$-values in high-accuracy settings do not necessarily indicate the absence of benefit, but rather motivate a closer examination of the $c/b$ counts to understand complementary effects.

For two strategies evaluated on the same set of problems (source-only baseline SC vs.\ Strategy~X), paired outcomes are summarized using the following $2\times2$ matrix:
\begin{center}
\scriptsize
\begin{tabular}{c|cc}
 & Strategy X \#correct & Strategy X \#wrong \\
\hline
Source-only \#correct & $a$ & $b$ \\
Source-only \#wrong   & $c$ & $d$
\end{tabular}
\end{center}
Here, $c$ (\emph{fixes}) denotes problems solved by Method~X but not by SC, while $b$ (\emph{regressions}) denotes problems solved by SC but not by Method~X.

McNemar's exact test is applied to the discordant pairs $(b,c)$. {Because we conduct multiple comparisons across language pairs and datasets, raw $p$-values are subject to Family-Wise Error Rate (FWER) inflation: running $k$ independent tests at $\alpha=0.05$ increases the probability of at least one false positive well above 5\%. To control for this, we apply the Holm--Bonferroni step-down correction~\citep{holm1979simple} within each dataset, treating all language-pair comparisons within a dataset as one family. The green/red shading in Table~\ref{tab:significance_result} reflects Holm-adjusted significance decisions at $\alpha=0.05$, not raw $p$-values.}

The results of these paired tests across datasets and language pairs are shown in Table~\ref{tab:significance_result}.
Overall, differences in aggregate success rates between Source-only (SC), \specS (NL), and \specS+ Source (NL + SC) are not statistically significant for most language pairs ($p \ge 0.05$), indicating that these approaches are often indistinguishable in terms of total accuracy alone.
However, relying solely on aggregate $p$-values can obscure the specific utility of the proposed methods.
The reported $c/b$ counts reveal a consistent presence of fixes across language pairs, demonstrating that \specS and combining \specS with source code successfully solve a subset of problems that the source-only strategy fails to handle.

{To explicitly capture this complementary behavior, we also evaluate  \emph{logical-OR ensemble} strategies, a problem is counted as solved if \emph{any} constituent strategy solves it. We note that the term ``OR'' here refers to a logical union of strategy outcomes, and is unrelated to the Odds Ratio statistic that McNemar's test can report; no inferential use of the Odds Ratio is made in this paper. These ensemble comparisons test whether combining strategies yields statistically significant improvements over the source-only baseline.}

{ Beyond statistical significance, we also report effect sizes using Cliff's 
$\delta$~\citep{cliff2014ordinal} for each pairwise comparison. Cliff's $\delta$ 
is non-parametric and well-suited to binary correct/incorrect outcomes. We interpret 
$|\delta|<0.147$ as negligible, $[0.147, 0.33)$ as small, $[0.33, 0.474)$ as medium, 
and $\geq0.474$ as large. Across language pairs, the observed effect sizes are 
predominantly negligible-to-small, indicating that even where statistical significance 
is achieved after Holm correction, the practical gains are modest. We therefore 
distinguish \textit{statistical} from \textit{practical} significance: \specS acts 
as a coverage-extending complementary signal rather than a large-magnitude accuracy 
improvement.}


{Taken together, these results show that \specS-based strategies act as complementary methods that broaden the coverage of source-only translation, rather than serving as strict replacements. Notably, no \specS-based strategy is significantly worse than source-only after Holm correction across any language pair or dataset, confirming that \specS is a safe complementary signal that extends coverage without significantly degrading accuracy. For instance, C$\rightarrow$Python under DeepSeek-Coder shows 13 regressions vs.\ 1 fix for the NL+SC strategy (raw $p<0.01$), yet this does not survive Holm correction within the CodeNet language pairs, illustrating the conservativeness of FWER control in large comparison families. The negligible-to-small Cliff's $\delta$ effect sizes across most language pairs further confirm that the gains are modest in magnitude, consistent with our conditional framing of \specS as a coverage-extending signal.}

\begin{table}

\centering
\tiny

\caption{Comparison of Critical and Blocker issues \emph{(count per 1\,000 NCLOC)} \textbf{before} fixing translated code across different models and settings.
S denotes using source code only,
N denotes using \specS\ only, and
N+S denotes using both source code and \specS.
For each row, the best result is highlighted.}
\label{sonar_table_models_before}
\begin{tabular}{|p{1cm}|p{.7cm}|p{.7cm}|p{.65cm}|p{.65cm}|p{.65cm}|p{.65cm}|p{.65cm}|p{.65cm}|p{.65cm}|p{.65cm}|p{.65cm}|}
\hline
Dataset & Source & Target 
& \multicolumn{3}{|c|}{GPT-4}
& \multicolumn{3}{|c|}{Magicoder}
& \multicolumn{3}{|c|}{DeepSee-Coderk} \\
& Lang & Lang 
& N & S & N+S
& N & S & N+S
& N & S & N+S \\
\hline
\multirow{8}{*}{\parbox{1.2cm}{\centering Avatar\\-\\Verified}} & Python & Java & 7.80 & 8.30 & 7.60 & 10.14 & 17.71 & 18.02 & 18.69 & \colorbox{highlight}{18.77} & 18.08 \\
 & Python & C & 43.20 & 45.90 & 44.10 & 45.10 & \colorbox{highlight}{50.10} & 50.08 & 36.41 & 34.63 & 35.02 \\
 & Python & C++ & 12.10 & 13.40 & 12.00 & 13.57 & \colorbox{highlight}{14.60} & 14.13 & 9.31 & 10.50 & 10.54 \\
 & Python & Go & 3.10 & 3.40 & 3.00 & 3.04 & 3.74 & 3.51 & 4.50 & \colorbox{highlight}{4.95} & 4.45 \\
 & Java & Python & 6.90 & 7.50 & 6.80 & \colorbox{highlight}{18.61} & 15.92 & 15.58 & 16.46 & 8.36 & 7.71 \\
 & Java & C & 42.80 & 44.50 & 41.90 & \colorbox{highlight}{46.33} & 45.80 & 43.97 & 30.42 & 29.00 & 31.69 \\
 & Java & C++ & 12.60 & 13.20 & 12.40 & \colorbox{highlight}{14.73} & 13.16 & - & 11.78 & 12.81 & 12.26 \\
 & Java & Go & 4.80 & 5.00 & 4.60 & - & 4.90 & - & \colorbox{highlight}{5.02} & 4.64 & 4.39 \\
\hline
\multirow{20}{*}{\textbf{CodeNet}} & Python & C & 44.90 & 47.20 & 43.80 & 44.17 & \colorbox{highlight}{50.88} & 49.97 & 47.43 & 41.33 & 39.14 \\
 & Python & C++ & 7.90 & 9.80 & 7.60 & 8.11 & \colorbox{highlight}{13.16} & 10.00 & 8.09 & 8.61 & 7.87 \\
& Python & Java & 13.60 & 15.40 & 13.20 & 12.83 & \colorbox{highlight}{18.44} & 18.09 & 18.28 & 14.89 & 16.56 \\
& Python & Go & 2.30 & 2.60 & 2.20 & 2.47 & 2.18 & 1.82 & 3.41 & 3.30 & \colorbox{highlight}{3.77} \\
& Java & C & 24.60 & 26.00 & 24.30 & 25.25 & 26.09 & 25.24 & 28.45 & \colorbox{highlight}{30.25} & 27.57 \\
& Java & C++ & 8.90 & 10.40 & 8.70 & 7.76 & 9.69 & 8.47 & 13.84 & \colorbox{highlight}{20.03} & 14.80 \\
& Java & Python & 14.90 & 16.00 & 14.20 & 16.17 & 14.87 & \colorbox{highlight}{22.11} & 9.11 & 6.51 & 6.15 \\
& Java & Go & 2.10 & 2.40 & 2.00 & 1.78 & 2.08 & 2.47 & 2.78 & 2.77 & \colorbox{highlight}{2.85} \\
& C & C++ & 41.90 & 44.80 & 41.50 & 42.71 & 45.17 & \colorbox{highlight}{51.84} & 45.27 & 48.86 & 43.93 \\
& C & Java & 11.30 & 12.40 & 11.00 & 11.71 & \colorbox{highlight}{13.96} & 13.11 & 10.33 & 11.29 & 12.16 \\
& C & Python & 12.80 & 14.40 & 12.30 & 13.50 & 15.10 & \colorbox{highlight}{16.69} & 13.16 & 12.42 & 9.51 \\
& C & Go & 2.70 & 3.00 & 2.60 & 2.38 & 3.45 & 2.91 & 3.84 & \colorbox{highlight}{3.94} & 3.81 \\
& C++ & C & 35.10 & 37.40 & 34.80 & 36.23 & 38.55 & \colorbox{highlight}{42.39} & 32.85 & 34.18 & 30.67 \\
& C++ & Java & 13.20 & 14.80 & 12.90 & 13.50 & \colorbox{highlight}{16.76} & 15.94 & 12.42 & 13.17 & 13.13 \\
& C++ & Python & 12.60 & 14.50 & 12.20 & 13.55 & \colorbox{highlight}{16.38} & 15.70 & 12.52 & 11.42 & 9.00 \\
& C++ & Go & 3.10 & 3.30 & 3.00 & 3.36 & 3.08 & 2.72 & 3.37 & 3.26 & \colorbox{highlight}{3.85} \\
& Go & C & 26.30 & 27.80 & 26.10 & 27.00 & 27.70 & 29.18 & 26.91 & \colorbox{highlight}{30.51} & 26.28 \\
& Go & C++ & 8.60 & 9.40 & 8.40 & 7.68 & 7.10 & 7.93 & 9.97 & \colorbox{highlight}{12.25} & 9.61 \\
& Go & Java & 16.90 & \colorbox{highlight}{18.10} & 16.40 & \colorbox{highlight}{18.10} & 16.47 & 17.92 & 8.32 & 7.11 & 5.98 \\
& Go & Python & 13.10 & 15.00 & 12.60 & 14.19 & 18.67 & \colorbox{highlight}{21.50} & 7.39 & 6.44 & 6.16 \\
\hline
EvalPlus & Python & Java & 48.90 & \colorbox{highlight}{5.20} & 4.90 & 2.13 & 3.33 & 1.95 & 2.12 & 2.12 & 1.96 \\
\hline
\end{tabular}
\end{table}

\begin{table}

\centering
\tiny
\caption{Comparison of Critical and Blocker issues \emph{(count per 1\,000 NCLOC)} \textbf{after} fixing translated code across different models and settings.
S denotes using source code only,
N denotes using \specS\ only, and
N+S denotes using both source code and \specS.
For each row, the best result is highlighted.}
\label{sonar_table_models_after}
\begin{tabular}{|p{1cm}|p{.7cm}|p{.7cm}|p{.65cm}|p{.65cm}|p{.65cm}|p{.65cm}|p{.65cm}|p{.65cm}|p{.65cm}|p{.65cm}|p{.65cm}|}
\hline
Dataset & Source & Target 
& \multicolumn{3}{|c|}{GPT-4}
& \multicolumn{3}{|c|}{Magicoder}
& \multicolumn{3}{|c|}{DeepSeek-Coder} \\
& Lang & Lang 
& N & S & N+S
& N & S & N+S
& N & S & N+S \\
\hline
\multirow{8}{*}{\parbox{1.2cm}{\centering Avatar\\-\\Verified}} & Java & C & 30.80 & \colorbox{highlight}{34.90} & 30.40 & 17.66 & 29.37 & 27.27 & 31.41 & 29.90 & 0 \\
& Java & C++ & 12.10 & 11.90 & 11.90 & \colorbox{highlight}{14.53} & 10.84 & - & 11.85 & 13.26 & 12.26 \\
& Java & Go & 4.70 & 5.00 & \colorbox{highlight}{5.10} & 4.96 & 4.59 & 4.57 & \colorbox{highlight}{5.10} & 4.92 & 4.29 \\
& Java & Python & 6.20 & 6.80 & 6.10 & \colorbox{highlight}{19.87} & 15.71 & 16.96 & 16.02 & 8.23 & 7.58 \\
& Python & C & \colorbox{highlight}{39.50} & 39.00 & 39.00 & 19.63 & - & 25.81 & 35.71 & 35.02 & 34.09 \\
& Python & C++ & 10.20 & 9.10 & 9.30 & 5.77 & 8.24 & 6.64 & 10.28 & \colorbox{highlight}{11.35} & 10.56 \\
& Python & Go & 4.10 & 4.40 & 4.80 & 2.00 & 3.59 & 3.39 & 5.04 & \colorbox{highlight}{5.14} & \colorbox{highlight}{5.14} \\
& Python & Java & 7.40 & 7.30 & 7.00 & 11.29 & \colorbox{highlight}{20.81} & 17.17 & 19.32 & 18.31 & 18.16 \\
\hline
\multirow{20}{*}{\textbf{CodeNet}} & C & C++ & 42.10 & 46.60 & 42.00 & 43.20 & 46.14 & \colorbox{highlight}{52.43} & 45.13 & 49.79 & 45.80 \\
& C & Go & 3.30 & 3.70 & 3.60 & 2.37 & 3.42 & 2.97 & 0 & \colorbox{highlight}{3.92} & 3.80 \\
& C & Java & 6.90 & 7.20 & 6.30 & 10.41 & \colorbox{highlight}{13.54} & 12.85 & 9.64 & 11.03 & 12.17 \\
& C & Python & 4.80 & 4.90 & 3.80 & 17.45 & 17.80 & \colorbox{highlight}{20.24} & 16.10 & 12.92 & 11.11 \\
& C++ & C & 35.00 & 35.20 & 34.20 & 34.59 & 38.87 & \colorbox{highlight}{43.14} & 33.21 & 33.83 & 31.36 \\
& C++ & Go & 3.60 & 4.00 & \colorbox{highlight}{4.50} & 3.51 & 3.06 & 2.72 & 3.66 & 3.27 & 3.99 \\
& C++ & Java & 7.60 & 8.60 & 7.40 & 11.97 & \colorbox{highlight}{15.85} & 15.27 & 13.02 & 13.44 & 13.24 \\
& C++ & Python & 3.10 & 6.30 & 5.20 & 16.82 & 16.57 & \colorbox{highlight}{17.24} & 15.42 & 11.83 & 11.29 \\
& Go & C & 32.00 & 32.70 & \colorbox{highlight}{34.20} & 27.15 & 29.46 & 31.23 & 24.60 & 27.84 & 24.70 \\
& Go & C++ & 11.00 & 11.10 & 10.60 & 7.04 & 6.99 & 7.59 & 10.36 & \colorbox{highlight}{12.00} & 9.67 \\
& Go & Java & 6.90 & 9.60 & 8.80 & 18.02 & \colorbox{highlight}{19.26} & 17.16 & 8.50 & 7.34 & 5.80 \\
& Go & Python & 5.00 & 5.70 & 3.80 & 14.62 & 19.47 & \colorbox{highlight}{23.75} & 7.88 & 6.31 & 6.31 \\
& Java & C & 31.50 & 32.80 & \colorbox{highlight}{34.90} & 24.98 & 26.25 & 24.69 & 28.95 & 30.04 & 27.67 \\
& Java & C++ & 13.00 & 12.90 & 12.50 & 8.25 & 10.11 & 8.61 & 14.08 & \colorbox{highlight}{20.39} & 15.16 \\
& Java & Go & 2.40 & 2.70 & 2.20 & 1.91 & 2.15 & 2.37 & 2.78 & 2.85 & \colorbox{highlight}{3.05} \\
& Java & Python & 4.10 & 5.30 & 3.90 & 17.35 & 14.84 & \colorbox{highlight}{21.21} & 10.77 & 7.08 & 7.22 \\
& Python & C & 49.20 & 42.50 & 44.50 & 44.41 & \colorbox{highlight}{50.99} & 48.79 & 46.10 & 41.88 & 39.19 \\
& Python & C++ & 10.00 & 6.60 & 6.60 & 8.14 & \colorbox{highlight}{13.14} & 9.89 & 0 & 8.60 & 7.87 \\
& Python & Go & 2.50 & 3.00 & 2.70 & 2.52 & 2.09 & 1.85 & 3.54 & 3.10 & \colorbox{highlight}{3.63} \\
& Python & Java & 4.80 & 4.30 & 4.70 & 12.31 & 17.18 & 16.93 & \colorbox{highlight}{17.91} & 14.28 & 16.02 \\
\hline
EvalPlus & Python & Java & \colorbox{highlight}{5.28} & 4.90 & 5.06 & 1.93 & 2.51 & 2.15 & 1.88 & 2.68 & 3.30 \\
\hline
\end{tabular}
\end{table}

\begin{figure}
    \centering
    \includegraphics[width=\textwidth]{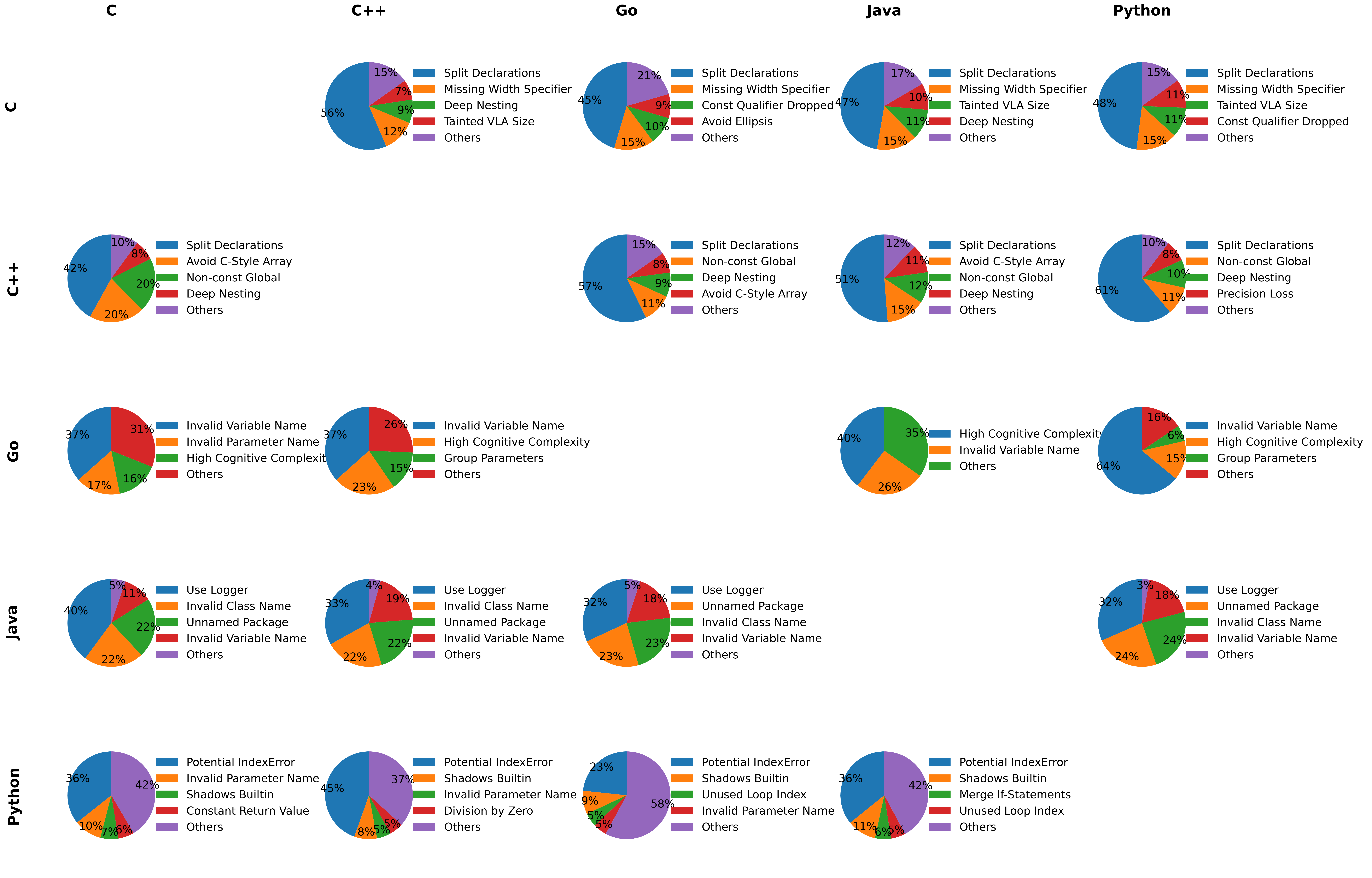} 
    \caption{Qualitative analysis of SonarQube warnings. Each pie chart represents the distribution of issue topics for a specific translation direction (Source $\rightarrow$ Target). The x-axis identifies the source languages, while the y-axis identifies the target languages.}
    \label{sonar_tags}
\end{figure}

\subsection*{\textbf{RQ3-\RQThree}}
{ \pa{Motivation.} Previous studies, such as Pan et al. (2024), have established baselines for code translation, but the impact of input modality on code hygiene is not well understood. We hypothesize that the nature of the input, specifically the use of high-level \specs versus low-level source code, might influence the LLM's adherence to language-specific best practices. RQ3 aims to quantify this impact. By analyzing the density of Critical and Blocker issues across our different experimental setups (\specS, source code, and \specS with source code), we assess whether explicit specifications help mitigate common static analysis warnings found in standard translation approaches.}

\pa{Method.}
To answer this research question, we use \citet{sonarqube} to analyze the quality of generated code both \emph{before} and \emph{after} repair. We evaluate three target models: DeepSeek-Coder, Magicoder-6.7B, and GPT-4, across all language pairs in the CodeNet, Avatar-Verified, and EvalPlus datasets. We first report the number of critical and blocker issues per 1000 NCLOC. We then present pie charts showing the distribution of issue categories aggregated over all language pairs. Details of the SonarQube experimental setup are described in Section~\ref{sonar_run}.

\noindent \textbf{Results.} {Tables~\ref{sonar_table_models_before} and~\ref{sonar_table_models_after} show detailed results before and after repair, respectively, for all three models across the three datasets and all language pairs. Both tables report the number of critical and blocker issues per 1000 NCLOC. Overall, the number of issues remains largely unchanged before and after repair, regardless of whether \specS only or \specS combined with the source code is used. Additionally, for a given dataset and language pair, the number of issues varies only slightly across different LLMs. However, some differences between LLMs are noticeable: for most language pairs, Magicoder tends to have a slightly higher number of issues compared to the other two LLMs, while GPT-4 generally has fewer issues. An exception is observed in the evalplus dataset, where GPT-4 produces more issues than the others. Another notable trend is that translations into the C language show a higher increase in issues compared to other language pairs, indicating that C code is more prone to vulnerabilities.}

\begin{boxFindings}
    Finding 3.1: The number of severe issues remains almost unchanged across approaches and between pre-fix and post-fix versions. 
\end{boxFindings}

\begin{boxFindings}
    Finding 3.2: With few exceptions, the source language has a negligible impact on the frequency and distribution of quality issues in the target language. Instead, the types of issues are primarily dictated by the target language's specific coding standards and paradigms.
\end{boxFindings}

{To analyze the underlying causes of the identified issues, we calculated the frequency of warnings for all source-to-target language pairs. We aggregated the SonarQube warning messages across all LLMs and categorized them into descriptive topics, which are visualized as pie charts in Figure \ref{sonar_tags}. 

As shown in the figure, the prevalent issues vary significantly depending on the target language. In C and C++, the most frequent issue is \textbf{Split Declaration}, which refers to defining multiple variables in a single statement (e.g., \texttt{int a, b;}) rather than using dedicated statements to improve readability and maintainability. In Go, the results are dominated by \textbf{Invalid Variable Names}, which violate the language's specific naming conventions (such as camelCase or underscore patterns), and \textbf{High Cognitive Complexity}, which indicates that the control flow is too difficult to follow due to deeply nested loops or conditionals. For Java, the primary issues are \textbf{Use Logger}, the recommendation to replace \texttt{System.out} with a logging framework, and naming violations for packages, classes, or variables. In Python, \textbf{Potential IndexError} is most common, indicating collection access that may exceed bounds due to mismatched indexing logic during translation. 

Generally, the source language does not significantly impact the frequency of issues in the target language. However, some exceptions exist: for instance, \textbf{Invalid Variable Names} are notably more frequent when translating from Python to Go, while \textbf{Split Declarations} are less common in C to C++ translations compared to other source languages.}


\noindent \textbf{Understanding the Spike in Defects for C Language Translations} As shown in Table~\ref{sonar_table_models_before} and \ref{sonar_table_models_after}, the number of quality issues typically remains below 10 per 1000 NLOC for most translation pairs. However, one notable exception emerges. Translations involving C (as either source or target) consistently result in significantly higher defect counts, often exceeding 30 per 1000 NLOC. To understand the primary causes of these widespread defects in C, we analyzed the most frequent SonarQube reports, which are listed below:
\begin{enumerate}
    \item Add a field width specifier to this "\%s" placeholder.. (\textbf{18.13\%})
    \item Refactor this code to not nest more than 3 if|for|do|while|switch statements. (\textbf{12.37\%})
    \item Declared variable-length array (VLA) has tainted (attacker-controlled) size that can be 0 or negative. (\textbf{10.69\%})
    \item cast from 'const void *' to 'int *' drops const qualifier. (\textbf{8.68\%})
    \item Replace this call to the non-reentrant function ``strtok'' by a call to ``strtok\_r''.. (\textbf{7.64\%})
    \item Division by a tainted value, possibly zero. (\textbf{4.99\%})
    \item Access of the heap area with a tainted index that may be negative or too large. (\textbf{2.46\%})
    \item Call to `malloc' has an allocation size of 0 bytes. (\textbf{0.97\%})
    \item call to undeclared library function `strtok' with type `char *(char *, const char *)'; ISO C99 and later do not support implicit function declarations. (\textbf{0.91\%})
    \item Access of `int' element in the heap area at index 1. (\textbf{0.78\%})
\end{enumerate}

\begin{boxFindings}
    Finding 3.3: The number of severe issues significantly increases when translating from C to other languages.
\end{boxFindings}

The top issue messages from SonarQube indicate several code quality and security issues in the C language. Key concerns include unsafe memory access (e.g., using tainted values for array indexing or division), improper use of functions like strtok (which should be replaced with strtok\_r for thread safety), and violations of C standards (e.g., implicit function declarations). This indicates that the language model is not highly concerned with the detailed and complex structure of the C language.

\section{Prescriptive Recommendations for Practitioners}
\label{sec:recommendations}

Based on our empirical findings across multiple datasets and language pairs, we offer the following operational guidelines for integrating LLM-based code translation into real-world workflows:

\begin{itemize}
    \item \textbf{Adopt a ``Source-First, Repair-Later'' Workflow.} 
    Since our statistical analysis shows that source-only translation achieves the highest general reliability, it should serve as the primary pass in any translation pipeline. However, rather than discarding failures, practitioners should implement a \textit{cascaded fallback}:
    \begin{enumerate}
        \item Attempt translation using the source-only baseline.
        \item If execution fails, trigger a second pass using the \textit{Source + \specS} approach.
    \end{enumerate}
    Our data confirms that \specS acts as a complementary signal, fixing a distinct subset of logic-heavy problems that the source-only model misses (the `fixes' denoted as $c$ in our results).

    \item \textbf{Target NL-Intermediates to High-Abstraction Languages.} 
    Practitioners should not apply NL-specification uniformly. Our analysis reveals that \specS yields the most benefit for \textbf{Python} and \textbf{C++} sources, where code is often dense with algorithmic logic. For these languages, the NL step acts as a ``denoising'' filter, separating intent from syntax. Conversely, for verbose languages like \textbf{Java} or system-level languages like \textbf{C}, the NL abstraction often discards critical low-level details (e.g., memory management, strict typing), leading to lower accuracy.

    \item \textbf{Inject Explicit Constraints for C-like Targets.} 
    When the target language requires manual memory management (e.g., C, C++), standard NL summaries are insufficient because they abstract away memory operations. Practitioners must modify their prompting strategy to explicitly request the preservation of memory allocation/deallocation logic when generating specifications for these targets, as standard prompts result in high compilation error rates due to missing headers or pointers.

    \item \textbf{Audit Test Suites for Logic vs. Syntax Errors.} 
    Reliable evaluation depends critically on the quality of the test suite. We observed that weak test cases often allow hallucinated logic to pass (false positives). Practitioners should prioritize ``mutation testing'' or high-coverage test suites when validating LLM-translated code, particularly when using NL-based methods, as the semantic drift in the intermediate NL step can introduce subtle boundary-condition bugs that simple I/O tests miss.
\end{itemize}

{
\section{Threats to validity}
This study acknowledges several threats to validity due to uncontrollable circumstances that may affect the dependability and applicability of our conclusions.

\subsection{Internal Validity.} Internal validity concerns the extent to which our results are attributable to our experimental design rather than confounding factors.
\begin{itemize}
    \item \textbf{Baseline Replication:} Our comparison relies on replicating the baseline study \citep{pan2024lost}. Although we validated results using the original artifacts with our cleaned and verified datapoints, minor inconsistencies in runtime environments or toolchains (e.g., compiler versions or dependency resolution) may affect compilation success, runtime behavior, or test execution, potentially leading to small variations, either positively or negatively, in pass/fail outcomes compared to the originally reported results.
    \item \textbf{Non-determinism of LLM Outputs:} LLM inference is inherently non-deterministic, and identical inputs may yield different outputs across runs. To mitigate this threat, we fix the model version, prompts, decoding parameters, and random seed wherever supported by the API, and execute each experiment using a single generation per datapoint, following the evaluation protocol of the baseline study \citep{pan2024lost}. This ensures consistent experimental conditions across all compared methods.
    \item \textbf{Prompt Sensitivity:} The prompts for NL-specification generation, code translation, compilation error repair were manually designed. While crafted carefully, LLMs are sensitive to prompt phrasing and structure, and alternative prompt variants could yield different results. To mitigate this, for NL-specification generation, we have conducted a prompt sensitivity study and chose pseudocode as the best performing natural language specification for code translation. For other prompts, we chose the same one as baseline \citep{pan2024lost}.
    \item \textbf{Temperature Randomness:} Although stochastic sampling remains even with a fixed temperature, we consistently use a temperature of 0.7 for all generations across all tasks and languages. By keeping this parameter constant for both our approach and the replicated baseline, we ensure that any randomness introduced by sampling affects all methods uniformly and does not bias the comparative analysis.
    \item \textbf{Stopping Conditions and Generation Length:} We employ consistent stopping criteria, including fixed maximum token limits and standardized termination conditions, across all experimental runs. This reduces variability due to premature truncation or excessively long outputs and ensures that all methods are evaluated under identical generation constraints.
    \item \textbf{API Version Drift:} Our experiments on gpt-4 and deepseek-coder rely on external LLM APIs whose underlying implementations may change over time. To reduce the impact of API version drift, we document the exact model identifiers, API versions, and inference parameters used in our experiments and conduct all evaluations within a bounded time window. While complete control over backend changes is not possible, these measures improve reproducibility and facilitate future comparisons.
    \item \textbf{SonarQube Target Files:} SonarQube analyzes only compilable files for compiled languages (C/C++ and Java) and only syntactically valid files for interpreted languages (Python and Go). Files that cannot be analyzed are excluded, which could introduce selection bias favoring certain methods or languages. To mitigate this, we evaluate all methods on the same compilable subset for each language, and we explicitly report per-language and per-method coverage. This limitation is inherent to static analysis tools requiring successful builds, and we acknowledge it as a potential internal validity threat.
\end{itemize}

\subsection{Construct Validity.} Construct validity concerns whether the chosen measurements accurately capture the intended concept, in our case, translation quality.
\begin{itemize}
\item \textbf{SonarQube Severity-Based Quality Metric:} We use the number of Blocker and Critical issues per 1,000 NCLOC reported by SonarQube as a proxy for translation quality. However, static analysis tools may conflate semantic defects with certain rule-specific or language-dependent heuristics, and some reported issues may not directly correspond to translation errors. To mitigate this threat, we intentionally exclude lower-severity issues (Major, Minor, Info), which are more likely to reflect stylistic or convention-related concerns, and focus on high-severity issues that predominantly capture correctness, security, or runtime-impacting defects. This choice aligns the metric more closely with translation correctness rather than code style.

\item \textbf{Rule Sets, Tool Versions, and Build Configuration:} SonarQube’s findings depend on the selected rule set, analyzer versions, and language-specific build configurations, particularly for compiled languages such as C/C++. To reduce construct variability, we use SonarQube via SonarCloud (free tier) with the default \textit{sonar-way} rule set, a fixed scanner version (sonar-scanner-cli-8.0.1.6346), and a uniform quality gate across all languages and datasets. Language-specific build requirements (e.g., bytecode-level analysis for Java and build-wrapper-based compilation context for C/C++) are explicitly standardized and documented. These measures ensure that differences in reported issues stem primarily from translation artifacts rather than analysis configuration choices.
\end{itemize}

\subsection{Conclusion Validity.}
Conclusion validity concerns whether the observed results reliably support our claims.
\begin{itemize}
\item \textbf{Statistical Testing and Variance:} While our study includes over 6,000 translations, only a subset was examined in depth for qualitative analysis (250 \specS translations and 100 selected mismatches). These datapoints were randomly sampled to reduce selection bias, and manual analysis was conducted independently by the first two authors, with disagreements resolved through discussion. Although the results may not fully generalize to large, multi-file projects, snippet-level evaluation is an established methodology that enables controlled measurement of translation correctness. Extending these approaches to repository-scale translation remains future work.
\end{itemize}

\subsection{External Validity.} External validity relates to the generalizability of our findings.
\begin{itemize}
    \item \textbf{Dataset Bias:} Our study is based on three datasets: Avatar, CodeNet, and EvalPlus. These datasets primarily consist of competitive programming problems, which may not represent the diversity and complexity of real-world production code (e.g., enterprise systems, embedded applications).
    \item \textbf{Programming Language Scope:} We focused on five widely-used languages (C, C++, Go, Java, Python). The findings may not generalize to other paradigms (e.g., functional languages like Haskell or OCaml) or niche languages (e.g., Kotlin, Swift).
    \item \textbf{Model Versioning:} All experiments were conducted using GPT-4, MagicCoder, and DeepSeek Coder as of their respective versions in late 2025. Future updates to these models or other LLMs may change translation behavior, potentially affecting reproducibility or generalizability.
    \item \textbf{Snippet-Level Translation:} Our study focuses on function-level (EvalPlus) or file-level (CodeNet, Avatar) translation, where each source file is translated and analyzed independently. Real-world projects often involve repository-level translation with cross-file dependencies, build systems, and packaging, which may influence translation behavior. Nevertheless, snippet-level evaluation provides a controlled setting to compare languages and datasets and to assess the core translation quality issues introduced by LLMs.
\end{itemize}


}

\section{Related Works}
%

There are also some rule-based transpilers available, such as C2Rust \citep{c2rustgithub}, CxGo \citep{gotranspilecxgo}. But these transpilers need a lot of human effort to correctly map every function, library method, and object for every pair of languages. The translations are also subject to low correctness.

\citet{bairi2024codeplan} present a task-agnostic, neuro-symbolic framework called CodePlan for repository-level code editing, aimed at solving complex software engineering tasks like package migration and temporal code edits.

Pan et. al.~\citep{pan2024lost} conduct a large-scale empirical study to evaluate the capabilities of general and code-specific LLM in automatic code translation between C, C++, Go, Java, and Python. Their analysis shows that the studies LLMs can successfully translate code ranging from 2.1\% to 47.3\%. They also propose a prompt crafting technique that can boost the performance up to 5.5\%. A similar approach, but with different datasets and leveraging the test case generation capability of LLM called UniTrans is proposed by \citet{yang2024exploring}.

\citet{yin2024rectifiercodetranslationcorrector} present Rectifier (model Code T5+), a fine-tuned model on the correct and repaired (manually or by LLM) code pairs to repair unsuccessful translations produced by LLM. \citet{pan2023stelocoderdecoderonlyllmmultilanguage} present a decoder-based LLM modifying StarCoder's \citep{li2023starcodersourceyou} model to achieve code translation from C++, C\#, JavaScript, Java, and PHP to Python.

To automate the translation task, \citet{yuan2024transagentllmbasedmultiagentcode} propose a multi-agent translation named TransAgent. The authors incorporate a multi-agent pipeline including test case generation, syntax error fixation, and semantic error fixation using code alignment. \citet{bhattarai2024enhancing} propose a Retrieval-Augmented Generation (RAG) based few shot learning technique.

\citet{nitin2024spectraenhancingcodetranslation} uses multi-modal specifications augmented with source code to translate using LLM. The specifications are - invariants, test cases, and descriptions. Specifications are generated by the LLM itself. At first, the LLM is asked to generate more than one candidate specification (invariants, descriptions). Then, using the method of self-consistency (using that specification to regenerate the source code) select the correct candidate
specification. Then using those candidates augmented with source code, LLM is asked to translate the code from C to Rust and C to Go. 

\citet{yu2023pseudocode} propose a two-stage model, AGL-Code, to tackle the pseudocode-to-code generation task, focusing on improving generation accuracy and repair efficiency. Their approach incorporates a Multi-Scale Pyramid Feature Extractor (MPFE) and a series of adaptive attention mechanisms, collectively known as AGLRepair. The MPFE component refines the encoder in the code generation phase by merging global and multi-scale local features, addressing challenges associated with long pseudocode sequences. In the repair phase, AGLRepair employs intra-line, inter-line, and bidirectional code-error message attention modules to capture contextual and diagnostic cues, enhancing error localization and correction. Experimental results on the SPoC \citep{kulal2019spoc} dataset demonstrate AGL-Code’s superior synthesis success rates, establishing it as a state-of-the-art model for both accurate pseudocode generation and efficient error repair.

Intertrans \citep{macedo2024intertrans} proposes a multi-hop code translation strategy that translates programs through one or more intermediate programming languages before reaching the target language. This approach exploits the observation that LLMs exhibit asymmetric strengths across language pairs (i.e., performing better on certain source–target combinations). While effective, INTERTRANS requires a large number of LLM calls, leading to high computational cost.

Beyond code translation, several recent studies investigate robustness, security, and bias in LLM-based code generation. These works analyze how LLMs behave under perturbed inputs \citep{rabbi2025multi}, generate secure code under adversarial or constrained settings \citep{li2024exploratory, li2025prompt, cheng2025cfceval}, and exhibit systematic biases in generated programs \citep{ling2025bias}. Although these studies focus on code generation rather than translation, they provide complementary insights into the reliability and failure modes of LLM-produced code, which are also relevant to understanding and improving code translation systems.

{
\section{Limitations}
\label{sec:limitations}
This study has several limitations that should be considered when interpreting the results.

First, the effectiveness of \specS depends on the quality of the generated specifications. In cases where specifications are incomplete, ambiguous, or overly abstract, the benefits of NL-specification may be reduced, and translation performance may degrade.

Second, we limit post-generation fixes primarily to compilation errors in order to isolate the impact of input representations. As a result, our reported performance may underestimate the achievable translation quality under more aggressive iterative refinement or human-in-the-loop settings.

Third, our analysis focuses on functional correctness as measured by test cases. Other aspects of translation quality, such as readability, maintainability, or stylistic alignment with target-language conventions, are not explicitly evaluated and remain directions for future work.

Fourth, while we use NL-specification as an intermediate representation, we do not explicitly evaluate bidirectional or round-trip translation (e.g., code $\rightarrow$ NL-spec $\rightarrow$ code) using automated back-translation metrics. Instead, we assess specification fidelity through targeted manual inspection (RQ2), which allows for fine-grained identification of semantic omissions but does not provide a quantitative measure of information loss. Evaluating round-trip translation at scale is left for future work and is explored in a concurrent study \citep{rabbi2025babelcoder}. 
}

{
\section{Ethics and Licensing Clarification}
\noindent This work uses and extends publicly available benchmark datasets, namely EvalPlus\citep{liu2024your}, CodeNet\citep{puri2021codenet}, and Avatar\citep{ahmad-etal-2021-avatar}, which are released under permissive licenses. We fully acknowledge and credit the original authors and sources of these datasets, and we explicitly document their licenses in the paper.

Our modifications are limited to non-semantic transformations required for cross-language evaluation and reproducibility. Specifically, we convert EvalPlus unit tests from Python to Java. These changes do not alter the original task definitions, problem statements, or evaluation intent.

We confirm that the redistribution of the modified test cases and solutions complies with the terms of the original licenses. Any applicable usage restrictions are preserved and clearly stated. No proprietary data, personal data, or sensitive information is involved in this work, and we do not identify or target individual developers or communities. Our goal is to improve benchmarking consistency and reproducibility in multilingual code generation research.
}

\section{Conclusion and Future Work}
In this work, we experiment with two approaches for translating source code from one programming language to another: one using \specS as an intermediate representation and providing it as a prompt to an LLM to generate code in the target language, and another combining \specS with the source code to generate target language code and also fix compilation errors using the LLM. { We found that the approach combining source code with \specS yields gains over the baseline on certain datasets and specific source languages, while the source-only baseline remains the strongest strategy overall when aggregated across all language pairs.} We analyzed the reasons behind the successes and failures of both approaches, as well as the quality of the translated code, identifying issues with the translation of C language code. In the future, we aim to enhance our approaches by addressing the root causes of failures and refining the intermediate representation. As future work, we also plan to extend our approach to repository-level code translation and improve performance in modest areas by incorporating Spectrum-Based Fault Localization (SBFL) techniques.

\section*{Acknowledgement}
\label{acknowledgement}
\noindent This research was supported by the Fonds de recherche du Québec (Grant No.2024-NOVA346499)\citep{nova}, Natural Sciences and Engineering Research Council of Canada (NSERC) through the Alliance, Grant (Grant No.586838-23), the NSERC Discovery Grant (Grant No. RGPIN-2019-07007 and Grant No. DGECR-2019-00464), and NSERC CREATE Grant (Grant No.555406-2021). We gratefully acknowledge the support of all funding agencies.
\section*{Declaration of generative AI and AI-assisted technologies in the manuscript preparation process}
\noindent During the preparation of this work, the author(s) used Claude and Grammarly in order to find related works and to polish the writing. After using this tool/service, the author(s) reviewed and edited the content as needed and take(s) full responsibility for the content of the published article.


\begin{thebibliography}{45}
\expandafter\ifx\csname natexlab\endcsname\relax\def\natexlab#1{#1}\fi
\providecommand{\url}[1]{\texttt{#1}}
\providecommand{\href}[2]{#2}
\providecommand{\path}[1]{#1}
\providecommand{\DOIprefix}{doi:}
\providecommand{\ArXivprefix}{arXiv:}
\providecommand{\URLprefix}{URL: }
\providecommand{\Pubmedprefix}{pmid:}
\providecommand{\doi}[1]{\href{http://dx.doi.org/#1}{\path{#1}}}
\providecommand{\Pubmed}[1]{\href{pmid:#1}{\path{#1}}}
\providecommand{\bibinfo}[2]{#2}
\ifx\xfnm\relax \def\xfnm[#1]{\unskip,\space#1}\fi
\bibitem[{Achiam et~al.(2023)Achiam, Adler, Agarwal, Ahmad, Akkaya, Aleman, Almeida, Altenschmidt, Altman, Anadkat et~al.}]{achiam2023gpt}
\bibinfo{author}{Achiam, J.}, \bibinfo{author}{Adler, S.}, \bibinfo{author}{Agarwal, S.}, \bibinfo{author}{Ahmad, L.}, \bibinfo{author}{Akkaya, I.}, \bibinfo{author}{Aleman, F.L.}, \bibinfo{author}{Almeida, D.}, \bibinfo{author}{Altenschmidt, J.}, \bibinfo{author}{Altman, S.}, \bibinfo{author}{Anadkat, S.}, et~al., \bibinfo{year}{2023}.
\newblock \bibinfo{title}{Gpt-4 technical report}.
\newblock \bibinfo{journal}{arXiv preprint arXiv:2303.08774} .
\bibitem[{Ahmad et~al.(2021)Ahmad, Tushar, Chakraborty and Chang}]{ahmad-etal-2021-avatar}
\bibinfo{author}{Ahmad, W.U.}, \bibinfo{author}{Tushar, M.G.R.}, \bibinfo{author}{Chakraborty, S.}, \bibinfo{author}{Chang, K.W.}, \bibinfo{year}{2021}.
\newblock \bibinfo{title}{Avatar: A parallel corpus for java-python program translation}.
\newblock \bibinfo{journal}{arXiv preprint arXiv:2108.11590} .
\bibitem[{Austin et~al.(2021)Austin, Odena, Nye, Bosma, Michalewski, Dohan, Jiang, Cai, Terry, Le and Sutton}]{austin2021programsynthesislargelanguage}
\bibinfo{author}{Austin, J.}, \bibinfo{author}{Odena, A.}, \bibinfo{author}{Nye, M.}, \bibinfo{author}{Bosma, M.}, \bibinfo{author}{Michalewski, H.}, \bibinfo{author}{Dohan, D.}, \bibinfo{author}{Jiang, E.}, \bibinfo{author}{Cai, C.}, \bibinfo{author}{Terry, M.}, \bibinfo{author}{Le, Q.}, \bibinfo{author}{Sutton, C.}, \bibinfo{year}{2021}.
\newblock \bibinfo{title}{Program synthesis with large language models}.
\newblock \URLprefix \url{https://arxiv.org/abs/2108.07732}.
\bibitem[{Bairi et~al.(2024)Bairi, Sonwane, Kanade, Iyer, Parthasarathy, Rajamani, Ashok and Shet}]{bairi2024codeplan}
\bibinfo{author}{Bairi, R.}, \bibinfo{author}{Sonwane, A.}, \bibinfo{author}{Kanade, A.}, \bibinfo{author}{Iyer, A.}, \bibinfo{author}{Parthasarathy, S.}, \bibinfo{author}{Rajamani, S.}, \bibinfo{author}{Ashok, B.}, \bibinfo{author}{Shet, S.}, \bibinfo{year}{2024}.
\newblock \bibinfo{title}{Codeplan: Repository-level coding using llms and planning}.
\newblock \bibinfo{journal}{Proceedings of the ACM on Software Engineering} \bibinfo{volume}{1}, \bibinfo{pages}{675--698}.
\bibitem[{Bhattarai et~al.(2024)Bhattarai, Santos, Jones, Biswas, Alexandrov and O'Malley}]{bhattarai2024enhancing}
\bibinfo{author}{Bhattarai, M.}, \bibinfo{author}{Santos, J.E.}, \bibinfo{author}{Jones, S.}, \bibinfo{author}{Biswas, A.}, \bibinfo{author}{Alexandrov, B.}, \bibinfo{author}{O'Malley, D.}, \bibinfo{year}{2024}.
\newblock \bibinfo{title}{Enhancing code translation in language models with few-shot learning via retrieval-augmented generation}.
\newblock \bibinfo{journal}{arXiv preprint arXiv:2407.19619} .
\bibitem[{Braun and Clarke(2006)}]{braun2006using}
\bibinfo{author}{Braun, V.}, \bibinfo{author}{Clarke, V.}, \bibinfo{year}{2006}.
\newblock \bibinfo{title}{Using thematic analysis in psychology}.
\newblock \bibinfo{journal}{Qualitative research in psychology} \bibinfo{volume}{3}, \bibinfo{pages}{77--101}.
\bibitem[{BV(2024)}]{tiobe}
\bibinfo{author}{BV, T.S.}, \bibinfo{year}{2024}.
\newblock \bibinfo{title}{Tiobe index for october 2024}.
\newblock \URLprefix \url{https://www.tiobe.com/tiobe-index/}.
\bibitem[{Chen et~al.(2021)Chen, Tworek, Jun, Yuan, Pinto, Kaplan, Edwards, Burda, Joseph, Brockman et~al.}]{chen2021evaluating}
\bibinfo{author}{Chen, M.}, \bibinfo{author}{Tworek, J.}, \bibinfo{author}{Jun, H.}, \bibinfo{author}{Yuan, Q.}, \bibinfo{author}{Pinto, H.P.D.O.}, \bibinfo{author}{Kaplan, J.}, \bibinfo{author}{Edwards, H.}, \bibinfo{author}{Burda, Y.}, \bibinfo{author}{Joseph, N.}, \bibinfo{author}{Brockman, G.}, et~al., \bibinfo{year}{2021}.
\newblock \bibinfo{title}{Evaluating large language models trained on code}.
\newblock \bibinfo{journal}{arXiv preprint arXiv:2107.03374} .
\bibitem[{Chen et~al.(2024)Chen, Fang and Monperrus}]{chen2024supersonic}
\bibinfo{author}{Chen, Z.}, \bibinfo{author}{Fang, S.}, \bibinfo{author}{Monperrus, M.}, \bibinfo{year}{2024}.
\newblock \bibinfo{title}{Supersonic: Learning to generate source code optimizations in c/c++}.
\newblock \bibinfo{journal}{IEEE Transactions on Software Engineering} .
\bibitem[{Cheng and Yang(2025)}]{cheng2025cfceval}
\bibinfo{author}{Cheng, C.}, \bibinfo{author}{Yang, J.}, \bibinfo{year}{2025}.
\newblock \bibinfo{title}{Cfceval: Evaluating security aspects in code generated by large language models}, in: \bibinfo{booktitle}{2025 2nd IEEE/ACM International Conference on AI-powered Software (AIware)}, \bibinfo{organization}{IEEE}. pp. \bibinfo{pages}{01--10}.
\bibitem[{Cliff(2014)}]{cliff2014ordinal}
\bibinfo{author}{Cliff, N.}, \bibinfo{year}{2014}.
\newblock \bibinfo{title}{Ordinal methods for behavioral data analysis}.
\newblock \bibinfo{publisher}{Psychology Press}.
\bibitem[{Cohen(1960)}]{cohen1960coefficient}
\bibinfo{author}{Cohen, J.}, \bibinfo{year}{1960}.
\newblock \bibinfo{title}{A coefficient of agreement for nominal scales}.
\newblock \bibinfo{journal}{Educational and psychological measurement} \bibinfo{volume}{20}, \bibinfo{pages}{37--46}.
\bibitem[{Dubey et~al.(2024)Dubey, Jauhri, Pandey, Kadian, Al-Dahle, Letman, Mathur, Schelten, Yang, Fan et~al.}]{dubey2024llama}
\bibinfo{author}{Dubey, A.}, \bibinfo{author}{Jauhri, A.}, \bibinfo{author}{Pandey, A.}, \bibinfo{author}{Kadian, A.}, \bibinfo{author}{Al-Dahle, A.}, \bibinfo{author}{Letman, A.}, \bibinfo{author}{Mathur, A.}, \bibinfo{author}{Schelten, A.}, \bibinfo{author}{Yang, A.}, \bibinfo{author}{Fan, A.}, et~al., \bibinfo{year}{2024}.
\newblock \bibinfo{title}{The llama 3 herd of models}.
\newblock \bibinfo{journal}{arXiv preprint arXiv:2407.21783} .
\bibitem[{{Fonds de recherche du Qu{\'e}bec}(2024)}]{nova}
\bibinfo{author}{{Fonds de recherche du Qu{\'e}bec}}, \bibinfo{year}{2024}.
\newblock \bibinfo{title}{{FRQNT-NSERC NOVA Program, Grant No. 2024-NOVA-346499}}.
\newblock \URLprefix \url{https://doi.org/10.69777/346499}, \DOIprefix\doi{10.69777/346499}.
\bibitem[{GoTranspile(n.d.)}]{gotranspilecxgo}
\bibinfo{author}{GoTranspile}, \bibinfo{year}{n.d.}
\newblock \bibinfo{title}{cxgo: Go to c++ transpiler}.
\newblock \URLprefix \url{https://github.com/gotranspile/cxgo}. \bibinfo{note}{accessed: [date of access]}.
\bibitem[{Hindle et~al.(2016)Hindle, Barr, Gabel, Su and Devanbu}]{hindle2016naturalness}
\bibinfo{author}{Hindle, A.}, \bibinfo{author}{Barr, E.T.}, \bibinfo{author}{Gabel, M.}, \bibinfo{author}{Su, Z.}, \bibinfo{author}{Devanbu, P.}, \bibinfo{year}{2016}.
\newblock \bibinfo{title}{On the naturalness of software}.
\newblock \bibinfo{journal}{Communications of the ACM} \bibinfo{volume}{59}, \bibinfo{pages}{122--131}.
\bibitem[{Holm(1979)}]{holm1979simple}
\bibinfo{author}{Holm, S.}, \bibinfo{year}{1979}.
\newblock \bibinfo{title}{A simple sequentially rejective multiple test procedure}.
\newblock \bibinfo{journal}{Scandinavian journal of statistics} , \bibinfo{pages}{65--70}.
\bibitem[{Immunant(2024)}]{c2rustgithub}
\bibinfo{author}{Immunant, I.}, \bibinfo{year}{2024}.
\newblock \bibinfo{title}{C2rust: Tools for translating c to rust}.
\newblock \URLprefix \url{https://github.com/immunant/c2rust}. \bibinfo{note}{accessed: 2024-11-12}.
\bibitem[{Kulal et~al.(2019)Kulal, Pasupat, Chandra, Lee, Padon, Aiken and Liang}]{kulal2019spoc}
\bibinfo{author}{Kulal, S.}, \bibinfo{author}{Pasupat, P.}, \bibinfo{author}{Chandra, K.}, \bibinfo{author}{Lee, M.}, \bibinfo{author}{Padon, O.}, \bibinfo{author}{Aiken, A.}, \bibinfo{author}{Liang, P.S.}, \bibinfo{year}{2019}.
\newblock \bibinfo{title}{Spoc: Search-based pseudocode to code}.
\newblock \bibinfo{journal}{Advances in Neural Information Processing Systems} \bibinfo{volume}{32}.
\bibitem[{Lachaux et~al.(2020)Lachaux, Roziere, Chanussot and Lample}]{lachaux2020unsupervised}
\bibinfo{author}{Lachaux, M.A.}, \bibinfo{author}{Roziere, B.}, \bibinfo{author}{Chanussot, L.}, \bibinfo{author}{Lample, G.}, \bibinfo{year}{2020}.
\newblock \bibinfo{title}{Unsupervised translation of programming languages}.
\newblock \bibinfo{journal}{arXiv preprint arXiv:2006.03511} .
\bibitem[{Landis and Koch(1977)}]{landis1977measurement}
\bibinfo{author}{Landis, J.R.}, \bibinfo{author}{Koch, G.G.}, \bibinfo{year}{1977}.
\newblock \bibinfo{title}{The measurement of observer agreement for categorical data}.
\newblock \bibinfo{journal}{biometrics} , \bibinfo{pages}{159--174}.
\bibitem[{Li et~al.(2024)Li, Rabbi, Cheng, Sangalay, Tian and Yang}]{li2024exploratory}
\bibinfo{author}{Li, J.}, \bibinfo{author}{Rabbi, F.}, \bibinfo{author}{Cheng, C.}, \bibinfo{author}{Sangalay, A.}, \bibinfo{author}{Tian, Y.}, \bibinfo{author}{Yang, J.}, \bibinfo{year}{2024}.
\newblock \bibinfo{title}{An exploratory study on fine-tuning large language models for secure code generation}.
\newblock \bibinfo{journal}{arXiv preprint arXiv:2408.09078} .
\bibitem[{Li et~al.(2025)Li, Rabbi, Yang, Wang and Yang}]{li2025prompt}
\bibinfo{author}{Li, J.}, \bibinfo{author}{Rabbi, F.}, \bibinfo{author}{Yang, B.}, \bibinfo{author}{Wang, S.}, \bibinfo{author}{Yang, J.}, \bibinfo{year}{2025}.
\newblock \bibinfo{title}{Prompt, synthesize, fine-tune: A secure code generation recipe}.
\newblock \bibinfo{journal}{arXiv e-prints} , \bibinfo{pages}{arXiv--2510}.
\bibitem[{Li et~al.(2023)Li, Allal, Zi, Muennighoff, Kocetkov, Mou, Marone, Akiki, Li, Chim, Liu, Zheltonozhskii, Zhuo, Wang, Dehaene, Davaadorj, Lamy-Poirier, Monteiro, Shliazhko, Gontier, Meade, Zebaze, Yee, Umapathi, Zhu, Lipkin, Oblokulov, Wang, Murthy, Stillerman, Patel, Abulkhanov, Zocca, Dey, Zhang, Fahmy, Bhattacharyya, Yu, Singh, Luccioni, Villegas, Kunakov, Zhdanov, Romero, Lee, Timor, Ding, Schlesinger, Schoelkopf, Ebert, Dao, Mishra, Gu, Robinson, Anderson, Dolan-Gavitt, Contractor, Reddy, Fried, Bahdanau, Jernite, Ferrandis, Hughes, Wolf, Guha, von Werra and de~Vries}]{li2023starcodersourceyou}
\bibinfo{author}{Li, R.}, \bibinfo{author}{Allal, L.B.}, \bibinfo{author}{Zi, Y.}, \bibinfo{author}{Muennighoff, N.}, \bibinfo{author}{Kocetkov, D.}, \bibinfo{author}{Mou, C.}, \bibinfo{author}{Marone, M.}, \bibinfo{author}{Akiki, C.}, \bibinfo{author}{Li, J.}, \bibinfo{author}{Chim, J.}, \bibinfo{author}{Liu, Q.}, \bibinfo{author}{Zheltonozhskii, E.}, \bibinfo{author}{Zhuo, T.Y.}, \bibinfo{author}{Wang, T.}, \bibinfo{author}{Dehaene, O.}, \bibinfo{author}{Davaadorj, M.}, \bibinfo{author}{Lamy-Poirier, J.}, \bibinfo{author}{Monteiro, J.}, \bibinfo{author}{Shliazhko, O.}, \bibinfo{author}{Gontier, N.}, \bibinfo{author}{Meade, N.}, \bibinfo{author}{Zebaze, A.}, \bibinfo{author}{Yee, M.H.}, \bibinfo{author}{Umapathi, L.K.}, \bibinfo{author}{Zhu, J.}, \bibinfo{author}{Lipkin, B.}, \bibinfo{author}{Oblokulov, M.}, \bibinfo{author}{Wang, Z.}, \bibinfo{author}{Murthy, R.}, \bibinfo{author}{Stillerman, J.}, \bibinfo{author}{Patel, S.S.}, \bibinfo{author}{Abulkhanov, D.}, \bibinfo{author}{Zocca, M.},
  \bibinfo{author}{Dey, M.}, \bibinfo{author}{Zhang, Z.}, \bibinfo{author}{Fahmy, N.}, \bibinfo{author}{Bhattacharyya, U.}, \bibinfo{author}{Yu, W.}, \bibinfo{author}{Singh, S.}, \bibinfo{author}{Luccioni, S.}, \bibinfo{author}{Villegas, P.}, \bibinfo{author}{Kunakov, M.}, \bibinfo{author}{Zhdanov, F.}, \bibinfo{author}{Romero, M.}, \bibinfo{author}{Lee, T.}, \bibinfo{author}{Timor, N.}, \bibinfo{author}{Ding, J.}, \bibinfo{author}{Schlesinger, C.}, \bibinfo{author}{Schoelkopf, H.}, \bibinfo{author}{Ebert, J.}, \bibinfo{author}{Dao, T.}, \bibinfo{author}{Mishra, M.}, \bibinfo{author}{Gu, A.}, \bibinfo{author}{Robinson, J.}, \bibinfo{author}{Anderson, C.J.}, \bibinfo{author}{Dolan-Gavitt, B.}, \bibinfo{author}{Contractor, D.}, \bibinfo{author}{Reddy, S.}, \bibinfo{author}{Fried, D.}, \bibinfo{author}{Bahdanau, D.}, \bibinfo{author}{Jernite, Y.}, \bibinfo{author}{Ferrandis, C.M.}, \bibinfo{author}{Hughes, S.}, \bibinfo{author}{Wolf, T.}, \bibinfo{author}{Guha, A.}, \bibinfo{author}{von Werra, L.},
  \bibinfo{author}{de~Vries, H.}, \bibinfo{year}{2023}.
\newblock \bibinfo{title}{Starcoder: may the source be with you!}
\newblock \URLprefix \url{https://arxiv.org/abs/2305.06161}.
\bibitem[{Ling et~al.(2025)Ling, Rabbi, Wang and Yang}]{ling2025bias}
\bibinfo{author}{Ling, L.}, \bibinfo{author}{Rabbi, F.}, \bibinfo{author}{Wang, S.}, \bibinfo{author}{Yang, J.}, \bibinfo{year}{2025}.
\newblock \bibinfo{title}{Bias unveiled: Investigating social bias in llm-generated code}, in: \bibinfo{booktitle}{Proceedings of the AAAI Conference on Artificial Intelligence}, pp. \bibinfo{pages}{27491--27499}.
\bibitem[{Liu et~al.(2024)Liu, Xia, Wang and Zhang}]{liu2024your}
\bibinfo{author}{Liu, J.}, \bibinfo{author}{Xia, C.S.}, \bibinfo{author}{Wang, Y.}, \bibinfo{author}{Zhang, L.}, \bibinfo{year}{2024}.
\newblock \bibinfo{title}{Is your code generated by chatgpt really correct? rigorous evaluation of large language models for code generation}.
\newblock \bibinfo{journal}{Advances in Neural Information Processing Systems} \bibinfo{volume}{36}.
\bibitem[{Lozhkov et~al.(2024)Lozhkov, Li, Allal, Cassano, Lamy-Poirier, Tazi, Tang, Pykhtar, Liu, Wei et~al.}]{lozhkov2024starcoder}
\bibinfo{author}{Lozhkov, A.}, \bibinfo{author}{Li, R.}, \bibinfo{author}{Allal, L.B.}, \bibinfo{author}{Cassano, F.}, \bibinfo{author}{Lamy-Poirier, J.}, \bibinfo{author}{Tazi, N.}, \bibinfo{author}{Tang, A.}, \bibinfo{author}{Pykhtar, D.}, \bibinfo{author}{Liu, J.}, \bibinfo{author}{Wei, Y.}, et~al., \bibinfo{year}{2024}.
\newblock \bibinfo{title}{Starcoder 2 and the stack v2: The next generation}.
\newblock \bibinfo{journal}{arXiv preprint arXiv:2402.19173} .
\bibitem[{Macedo et~al.(2024)Macedo, Tian, Nie, Cogo and Adams}]{macedo2024intertrans}
\bibinfo{author}{Macedo, M.}, \bibinfo{author}{Tian, Y.}, \bibinfo{author}{Nie, P.}, \bibinfo{author}{Cogo, F.R.}, \bibinfo{author}{Adams, B.}, \bibinfo{year}{2024}.
\newblock \bibinfo{title}{Intertrans: Leveraging transitive intermediate translations to enhance llm-based code translation}.
\newblock \bibinfo{journal}{arXiv preprint arXiv:2411.01063} .
\bibitem[{Nijkamp et~al.(2022)Nijkamp, Pang, Hayashi, Tu, Wang, Zhou, Savarese and Xiong}]{nijkamp2022codegen}
\bibinfo{author}{Nijkamp, E.}, \bibinfo{author}{Pang, B.}, \bibinfo{author}{Hayashi, H.}, \bibinfo{author}{Tu, L.}, \bibinfo{author}{Wang, H.}, \bibinfo{author}{Zhou, Y.}, \bibinfo{author}{Savarese, S.}, \bibinfo{author}{Xiong, C.}, \bibinfo{year}{2022}.
\newblock \bibinfo{title}{Codegen: An open large language model for code with multi-turn program synthesis}.
\newblock \bibinfo{journal}{arXiv preprint arXiv:2203.13474} .
\bibitem[{Nitin et~al.(2024)Nitin, Krishna and Ray}]{nitin2024spectraenhancingcodetranslation}
\bibinfo{author}{Nitin, V.}, \bibinfo{author}{Krishna, R.}, \bibinfo{author}{Ray, B.}, \bibinfo{year}{2024}.
\newblock \bibinfo{title}{Spectra: Enhancing the code translation ability of language models by generating multi-modal specifications}.
\newblock \URLprefix \url{https://arxiv.org/abs/2405.18574}.
\bibitem[{Pan et~al.(2023)Pan, Sadé, Kim, Soriano, Sole and Flamant}]{pan2023stelocoderdecoderonlyllmmultilanguage}
\bibinfo{author}{Pan, J.}, \bibinfo{author}{Sadé, A.}, \bibinfo{author}{Kim, J.}, \bibinfo{author}{Soriano, E.}, \bibinfo{author}{Sole, G.}, \bibinfo{author}{Flamant, S.}, \bibinfo{year}{2023}.
\newblock \bibinfo{title}{Stelocoder: a decoder-only llm for multi-language to python code translation}.
\newblock \URLprefix \url{https://arxiv.org/abs/2310.15539}.
\bibitem[{Pan et~al.(2024)Pan, Ibrahimzada, Krishna, Sankar, Wassi, Merler, Sobolev, Pavuluri, Sinha and Jabbarvand}]{pan2024lost}
\bibinfo{author}{Pan, R.}, \bibinfo{author}{Ibrahimzada, A.R.}, \bibinfo{author}{Krishna, R.}, \bibinfo{author}{Sankar, D.}, \bibinfo{author}{Wassi, L.P.}, \bibinfo{author}{Merler, M.}, \bibinfo{author}{Sobolev, B.}, \bibinfo{author}{Pavuluri, R.}, \bibinfo{author}{Sinha, S.}, \bibinfo{author}{Jabbarvand, R.}, \bibinfo{year}{2024}.
\newblock \bibinfo{title}{Lost in translation: A study of bugs introduced by large language models while translating code}, in: \bibinfo{booktitle}{Proceedings of the IEEE/ACM 46th International Conference on Software Engineering}, pp. \bibinfo{pages}{1--13}.
\bibitem[{Paul et~al.(2024)Paul, Glava{\v{s}} and Gurevych}]{paul2024ircoder}
\bibinfo{author}{Paul, I.}, \bibinfo{author}{Glava{\v{s}}, G.}, \bibinfo{author}{Gurevych, I.}, \bibinfo{year}{2024}.
\newblock \bibinfo{title}{Ircoder: Intermediate representations make language models robust multilingual code generators}.
\newblock \bibinfo{journal}{arXiv preprint arXiv:2403.03894} .
\bibitem[{Puri et~al.(2021)Puri, Kung, Janssen, Zhang, Domeniconi, Zolotov, Dolby, Chen, Choudhury, Decker et~al.}]{puri2021codenet}
\bibinfo{author}{Puri, R.}, \bibinfo{author}{Kung, D.S.}, \bibinfo{author}{Janssen, G.}, \bibinfo{author}{Zhang, W.}, \bibinfo{author}{Domeniconi, G.}, \bibinfo{author}{Zolotov, V.}, \bibinfo{author}{Dolby, J.}, \bibinfo{author}{Chen, J.}, \bibinfo{author}{Choudhury, M.}, \bibinfo{author}{Decker, L.}, et~al., \bibinfo{year}{2021}.
\newblock \bibinfo{title}{Codenet: A large-scale ai for code dataset for learning a diversity of coding tasks}.
\newblock \bibinfo{journal}{arXiv preprint arXiv:2105.12655} .
\bibitem[{Rabbi et~al.(2025a)Rabbi, Ding and Yang}]{rabbi2025multi}
\bibinfo{author}{Rabbi, F.}, \bibinfo{author}{Ding, Z.}, \bibinfo{author}{Yang, J.}, \bibinfo{year}{2025}a.
\newblock \bibinfo{title}{A multi-language perspective on the robustness of llm code generation}.
\newblock \bibinfo{journal}{arXiv preprint arXiv:2504.19108} .
\bibitem[{Rabbi et~al.(2025b)Rabbi, Saha, Pham, Wang and Yang}]{rabbi2025babelcoder}
\bibinfo{author}{Rabbi, F.}, \bibinfo{author}{Saha, S.K.}, \bibinfo{author}{Pham, T.M.T.}, \bibinfo{author}{Wang, S.}, \bibinfo{author}{Yang, J.}, \bibinfo{year}{2025}b.
\newblock \bibinfo{title}{Babelcoder: Agentic code translation with specification alignment}.
\newblock \bibinfo{journal}{arXiv preprint arXiv:2512.06902} .
\bibitem[{SonarQube(2025)}]{sonarqube}
\bibinfo{author}{SonarQube}, \bibinfo{year}{2025}.
\newblock \bibinfo{title}{Sonarqube static code analysis}.
\newblock \URLprefix \url{https://www.sonarsource.com/}.
\bibitem[{Szafraniec et~al.(2022)Szafraniec, Roziere, Leather, Charton, Labatut and Synnaeve}]{szafraniec2022code}
\bibinfo{author}{Szafraniec, M.}, \bibinfo{author}{Roziere, B.}, \bibinfo{author}{Leather, H.}, \bibinfo{author}{Charton, F.}, \bibinfo{author}{Labatut, P.}, \bibinfo{author}{Synnaeve, G.}, \bibinfo{year}{2022}.
\newblock \bibinfo{title}{Code translation with compiler representations}.
\newblock \bibinfo{journal}{arXiv preprint arXiv:2207.03578} .
\bibitem[{Wei et~al.(2023)Wei, Wang, Liu, Ding and Zhang}]{wei2023magicoder}
\bibinfo{author}{Wei, Y.}, \bibinfo{author}{Wang, Z.}, \bibinfo{author}{Liu, J.}, \bibinfo{author}{Ding, Y.}, \bibinfo{author}{Zhang, L.}, \bibinfo{year}{2023}.
\newblock \bibinfo{title}{Magicoder: Empowering code generation with oss-instruct}.
\newblock \bibinfo{journal}{arXiv preprint arXiv:2312.02120} .
\bibitem[{Yang et~al.(2024)Yang, Liu, Yu, Keung, Li, Liu, Hong, Ma, Jin and Li}]{yang2024exploring}
\bibinfo{author}{Yang, Z.}, \bibinfo{author}{Liu, F.}, \bibinfo{author}{Yu, Z.}, \bibinfo{author}{Keung, J.W.}, \bibinfo{author}{Li, J.}, \bibinfo{author}{Liu, S.}, \bibinfo{author}{Hong, Y.}, \bibinfo{author}{Ma, X.}, \bibinfo{author}{Jin, Z.}, \bibinfo{author}{Li, G.}, \bibinfo{year}{2024}.
\newblock \bibinfo{title}{Exploring and unleashing the power of large language models in automated code translation}.
\newblock \bibinfo{journal}{Proceedings of the ACM on Software Engineering} \bibinfo{volume}{1}, \bibinfo{pages}{1585--1608}.
\bibitem[{Yin et~al.(2024)Yin, Ni, Nguyen, Wang and Yang}]{yin2024rectifiercodetranslationcorrector}
\bibinfo{author}{Yin, X.}, \bibinfo{author}{Ni, C.}, \bibinfo{author}{Nguyen, T.N.}, \bibinfo{author}{Wang, S.}, \bibinfo{author}{Yang, X.}, \bibinfo{year}{2024}.
\newblock \bibinfo{title}{Rectifier: Code translation with corrector via llms}.
\newblock \URLprefix \url{https://arxiv.org/abs/2407.07472}.
\bibitem[{Yu et~al.(2023)Yu, Huang and Gu}]{yu2023pseudocode}
\bibinfo{author}{Yu, Q.}, \bibinfo{author}{Huang, Z.}, \bibinfo{author}{Gu, N.}, \bibinfo{year}{2023}.
\newblock \bibinfo{title}{Pseudocode to code based on adaptive global and local information}, in: \bibinfo{booktitle}{2023 IEEE International Conference on Software Analysis, Evolution and Reengineering (SANER)}, \bibinfo{organization}{IEEE}. pp. \bibinfo{pages}{61--72}.
\bibitem[{Yuan et~al.(2024)Yuan, Chen, Wang, Yu, Peng and Lou}]{yuan2024transagentllmbasedmultiagentcode}
\bibinfo{author}{Yuan, Z.}, \bibinfo{author}{Chen, W.}, \bibinfo{author}{Wang, H.}, \bibinfo{author}{Yu, K.}, \bibinfo{author}{Peng, X.}, \bibinfo{author}{Lou, Y.}, \bibinfo{year}{2024}.
\newblock \bibinfo{title}{Transagent: An llm-based multi-agent system for code translation}.
\newblock \URLprefix \url{https://arxiv.org/abs/2409.19894}.
\bibitem[{Zheng et~al.(2023)Zheng, Xia, Zou, Dong, Wang, Xue, Wang, Shen, Wang, Li et~al.}]{zheng2023codegeex}
\bibinfo{author}{Zheng, Q.}, \bibinfo{author}{Xia, X.}, \bibinfo{author}{Zou, X.}, \bibinfo{author}{Dong, Y.}, \bibinfo{author}{Wang, S.}, \bibinfo{author}{Xue, Y.}, \bibinfo{author}{Wang, Z.}, \bibinfo{author}{Shen, L.}, \bibinfo{author}{Wang, A.}, \bibinfo{author}{Li, Y.}, et~al., \bibinfo{year}{2023}.
\newblock \bibinfo{title}{Codegeex: A pre-trained model for code generation with multilingual evaluations on humaneval-x}.
\newblock \bibinfo{journal}{arXiv preprint arXiv:2303.17568} .
\bibitem[{Zhu et~al.(2024)Zhu, Guo, Shao, Yang, Wang, Xu, Wu, Li, Gao, Ma et~al.}]{zhu2024deepseek}
\bibinfo{author}{Zhu, Q.}, \bibinfo{author}{Guo, D.}, \bibinfo{author}{Shao, Z.}, \bibinfo{author}{Yang, D.}, \bibinfo{author}{Wang, P.}, \bibinfo{author}{Xu, R.}, \bibinfo{author}{Wu, Y.}, \bibinfo{author}{Li, Y.}, \bibinfo{author}{Gao, H.}, \bibinfo{author}{Ma, S.}, et~al., \bibinfo{year}{2024}.
\newblock \bibinfo{title}{Deepseek-coder-v2: Breaking the barrier of closed-source models in code intelligence}.
\newblock \bibinfo{journal}{arXiv preprint arXiv:2406.11931} .

\end{thebibliography}
\end{document}

\endinput